\documentclass[aps,prd,reprint,superscriptaddress,nofootinbib]{revtex4-2}

\usepackage{graphicx}
\usepackage{subcaption}
\usepackage{amssymb,amsmath, amsfonts}
\usepackage{adjustbox}
\usepackage{lipsum}
\usepackage{epsfig}
\usepackage{xcolor}
\usepackage{tabularx}
\usepackage{soul}
\usepackage{multirow}
\usepackage{lineno,hyperref}

\usepackage{inputenc}
\usepackage{booktabs}
\usepackage{float}

\usepackage{lmodern}

\newcommand{\apjl}{Astrophys.\ J.\ Lett.}
\newcommand{\apjs}{Astrophys.\ J.\ Suppl.}
\newcommand{\mnras}{Mon.\ Not.\ R.\ Astron.\ Soc.}
\newcommand{\araa}{Annu.\ Rev.\ Astron.\ Astrophys.}
\newcommand{\aap}{Astron.\ Astrophys.}

\newcommand{\pasp}{Publ.\ Astron.\ Soc.\ Pac.}
\newcommand{\apss}{Astrophys.\ Space Sci.}

\usepackage{microtype}
\DeclareUnicodeCharacter{2009}{\,}
\DeclareUnicodeCharacter{2212}{-}

\begin{document}

\title{\textbf{Comprehensive Variability Analysis of Blazars Using Fermi Light Curves Across Multiple Timescales.} }

\author{Zahir Shah}
\email{shahzahir4@gmail.com}
\affiliation{Department of Physics, Central University of Kashmir, Ganderbal-191131, India}

\author{Athar A. Dar}
\affiliation{Department of Physics, Central University of Kashmir, Ganderbal-191131, India}
\affiliation{Department of Physics, University of Kashmir,  Srinagar-190006, India}

\author{Sikandar Akbar}
\email{darprince46@gmail.com}
\affiliation{Department of Physics, University of Kashmir,  Srinagar-190006, India}

\author{Anjum Peer}
\affiliation{Department of Physics, Islamic University of Science and Technology, Awantipora-192122, India}

\author{Zahoor Malik}
\affiliation{Department of Physics, National Institute of Technology,  Srinagar-190006, India}

\author{Aaqib Manzoor}
\affiliation{Indian Institute of Astrophysics, Bangalore, India}

\author{Sajad Ahanger}
\affiliation{Department of Physics, University of Kashmir,  Srinagar-190006, India}

\author{Javaid Tantry}
\affiliation{Department of Physics, University of Kashmir,  Srinagar-190006, India}

\author{Zeeshan Nazir}
\affiliation{Department of Physics, Central University of Kashmir, Ganderbal-191131, India}

\author{Debanjan Bose}
\affiliation{Department of Physics, Central University of Kashmir, Ganderbal-191131, India}

\author{Mushtaq Magray}
\affiliation{Department of Physics, Central University of Kashmir, Ganderbal-191131, India}

\begin{abstract}
In this study, we conducted a systematic analysis of long-term \emph{Fermi}-LAT $\gamma$-ray data for a sample of blazars, including FSRQs,  BL\,Lacs, and BCUs, to investigate their $\gamma$-ray variability. We focused on light curves binned in 3-day, 7-day, and 30-day intervals to assess the impact of binning on variability, using data with $TS >4$ as a detection threshold. We calculated fractional variability ($\rm F_{var}$) for each catagory and found that FSRQs exhibit higher mean variability compared to BL\,Lacs and BCUs, with BCUs displaying intermediate variability closer to BL\,Lacs.   The Kolmogorov-Smirnov (KS) test on the variability distributions of FSRQs, BL Lacs, and BCUs indicates that FSRQs differ from both BL\,Lacs and BCUs, whereas BCUs are more similar to BL\,Lacs.
The observed higher variability in FSRQs is likely linked to more powerful jets and accretion processes. The correlation between $\gamma$-ray flux and spectral index suggest a moderate positive correlation for BL\,Lacs and BCUs, indicating a ``softer when brighter" behavior. FSRQs, however, displayed a mild anticorrelation, suggesting a tendency for these sources to become harder as their flux increases.  Additionally, the analysis of flux distributions revealed log-normal behavior in many sources, consistent with multiplicative variability processes in blazar jets. Some sources exhibit bimodal distributions, implying transitions between distinct emission states. Moreover, binning effects the observed variability, with longer bins smoothing short-term fluctuations.  The power spectral density (PSD) analysis suggested that FSRQs exhibit steeper slopes, reflecting structured variability, while BL\, Lacs display shallower slopes, dominated by stochastic processes. The absence of PSD breaks suggests no dominant characteristic timescale within the \emph{Fermi} window. Spectral index distributions further highlight complexity, often requiring multi-component models. 
\end{abstract}

\maketitle

\section{Introduction}
\label{introduction}
The {\emph Fermi} Large Area Telescope \cite{2009ApJ...697.1071A} has revolutionized our understanding of the $\gamma$-ray variable sky. With its large field of view (covering 20\% of the sky) and its ability to scan the entire sky in approximately 3 hours, the LAT offers enhanced effective area and sensitivity. This continuous monitoring capability has made it an essential observatory for temporal and spectral studies in $\gamma$ rays. Over more than ten years, the LAT has been instrumental in identifying and consistently tracking numerous variable celestial sources \citep{2022ApJS..260...53A}.

Blazars are the dominant extragalactic sources detected in the $\gamma$-ray sky by the {\emph Fermi}-LAT  \citep{2010ApJS..188..405A}. These are a subclass of active galactic nuclei (AGNs) distinguished by their jets, which are directed towards our line of sight \citep{1995PASP..107..803U}. Blazars are notable for their intense $\gamma$-ray emissions, produced by relativistic jets. These jets, composed of non-thermal plasma, originate near supermassive black holes and extend over kpc/Mpc scales  \citep{2019ARA&A..57..467B}.  Blazars exhibit variability in both flux and spectrum across the electromagnetic spectrum, with timescales ranging from minutes to years \citep{2007ApJ...666L..17A,2016ApJ...824L..20A}. Their high-energy emission and broadband variability (radio-to-$\gamma$-ray) make them crucial targets for multiwavelength studies  \citep{1997ARA&A..35..445U,2006ASPC..350.....M}. The continuous monitoring capabilities of {\emph Fermi}-LAT in the 20 MeV to over 300 GeV range makes it an essential tool for investigating blazar variability in high energy regime \citep{2009ApJ...697.1071A,2009ApJ...700..597A,2010ApJ...716...30A}.

Blazars are commonly divided into two groups: BL\,Lac objects and Flat Spectrum Radio Quasars (FSRQs), primarily distinguished by their optical spectrum characteristics. BL Lac objects exhibit weak or absent emission line features (EW $<$ 5 Å), while FSRQs display prominent emission lines (EW $>$ 5 \AA) \citep{1995PASP..107..803U}. The broadband SED of blazars shows two humps, the first hump peaks in the optical/UV/X-ray energies, while the second hump peaks in the GeV energies. The low energy hump is well explained by the synchrotron emission from the relativistic electrons, while the high energy hump is mostly explained by the inverse Compton scattering (IC)  \citep{2013ApJ...768...54B, 2017MNRAS.470.3283S, 2021MNRAS.504..416S, 2024MNRAS.527.5140S}.  Alternatively, high-energy emission can also arise from hadronic processes, where relativistic protons drive emission through proton-synchrotron radiation and pion production mechanisms  \citep{2001APh....15..121M, 2013ApJ...768...54B}.

 Since the launch of the {\emph Fermi} $\gamma$-ray  telescope in 2008, the number of blazars that are known to be emitters of $\gamma$-rays has drastically increased. The study of $\gamma$-ray variability in blazars has become a major focus in astrophysics.  The long-term monitoring provided by the LAT has been essential for understanding the variability patterns of blazars, shedding light on the underlying physical processes at play. The continuous observations by the LAT have enabled extensive studies of blazar variability across multiple timescales, from minutes to years. This has been particularly valuable in multi-wavelength campaigns that aim to study long-term correlated variability in AGN \citep{10.1093/mnras/stab3358}. For instance, the detection of systematic lags between optical and $\gamma$-ray flares \citep{2012ApJ...754..114H,2015ApJ...807...79H} and evidence of quasi-periodic variations in BL Lac objects \citep{2015ApJ...813L..41A} have provided critical insights into the physics of relativistic jets.

Despite extensive studies over the years, most blazar research has focused on individual sources using multi-wavelength data \citep{2009ApJ...697L..81B, 2019MNRAS.484.3168S, 2024ApJ...977..111A, 2024JHEAp..44..393T}. However, few studies have examined the $\gamma$-ray flux variability characteristics of a large sample of blazars. Notable early work by \cite{2010ApJS..188..405A} analyzed 11 months of {\emph Fermi}-LAT data for 106 objects, and more recent research by \cite{2018ApJ...853..159L} investigated the $\gamma$-ray flux variability of high-redshift (z $>$ 3) blazars. Systematic studies of $\gamma$-ray variability in large blazar samples using {\emph Fermi}-LAT data are crucial for advancing our understanding of these enigmatic sources. By focusing on $\gamma$-ray flux variability, we can uncover patterns and behaviors that are key to deciphering the physical processes driving blazar emissions.

The long-term data set available for a large sample of blazars allows for comprehensive analyses to characterize their $\gamma$-ray variability. This study aims to characterize the long-term $\gamma$-ray variability of blazars on different time scales (3-day, 7-day, and 30-day). The characterization includes examining flux variability amplitude, flux distribution, index-flux relation, power spectral density (PSD), and spectral distribution. We utilize light curves from the {\emph Fermi}-LAT Light Curve Repository (LCR), developed by \citet{2023ApJS..265...31A}. This public database provides light curves for various {\emph Fermi}-LAT sources across different time scales, produced through likelihood analyses of the sources and their surroundings, yielding measurements of flux and spectral index. The paper is organized as follows: Section \S 2 gives the details of the sample of blazars; Section \S 3 discusses the temporal study of blazars, including spectral index distribution and flux-index correlation; Sections \S 4 and \S 5 cover the PSD and index distribution study, respectively; Section \S 6 offers a summary and discussion; and finally, the conclusion is provided in Section \S 7.

\section{Sample of Blazars } \label{sec:data_ana}
\subsection{\emph{Fermi}-LAT}
The {\emph Fermi}-LAT is a pair-conversion $\gamma$-ray telescope designed to detect photons with energies exceeding 20 MeV \citep{2009ApJ...697.1071A,2009ApJ...707.1310A}.   The LAT operates mainly in a scanning mode, surveying the entire sky every three hours \citep{2009ApJ...697.1071A}. It has a large effective area of around 8000 $cm^2$ for 1 GeV photons and has a broad field of view of approximately  2.4 steradians. This research focuses on extracting $\gamma$-ray data from the {\emph Fermi}-LCR for all recorded blazars.

The LCR targets sources listed in the 4FGL-DR2 catalog with variability indices greater than 21.67 \citep{2020arXiv200511208B,Abdollahi_2023}. This catalog, based on a 10-year survey, indicates that sources with a variability index above 21.67 over 12 intervals have less than a 1\% chance of remaining steady. Most of these variable sources are blazars, categorized as flat spectrum radio quasars (FSRQ), BL Lacs, and blazar candidates of unknown type (BCU). The LCR provide 3-day, 7-day, and 30-day binned light curves for each of these sources over more than 13 years of data involves analyzing.  The LCR includes 794 sources as FSRQs, 1456 sources as BL\,Lacs and 1493 sources as BCUs.
The LCR analysis employs the standard {\emph Fermi}-LAT science tool (version v11r5p3) and utilizes the $P8R2_-SOURCE_-V6$ instrument response functions on $P8R3_-SOURCE$ class photons within an energy range of 100 MeV to 100 GeV. Light curves generated by the LCR are derived through unbinned likelihood analysis. The count distribution for each source of interest is modeled as a point source, incorporating all $\gamma$-ray sources from the 4FGL-DR2 catalog within a 30° radius of each region of interest (ROI). The normalization of each variable source in the region is allowed to vary, while the spectral parameters remain fixed to their catalog values. The model also includes Galactic and isotropic background components. The Galactic component $gll_-iem_-v07.fits$ accounts for interstellar diffuse $\gamma$-ray emission from the Milky Way, and the isotropic component $iso_-P8R3_-SOURCE_-V3_-v1$ accounts for residual charged-particle backgrounds and isotropic celestial $\gamma$-ray emission. Both Galactic and isotropic component normalizations are set free to vary during the fit.

To maximize the likelihood of the observed data given the model, the free parameters are adjusted. Initially, the normalization of the source spectrum varies freely, while spectral parameters are fixed to their 4FGL-DR2 catalog values. After achieving a satisfactory fit, a second fitting round allows the photon index of the source of interest to vary. The likelihood analysis uses a test statistic, which is twice the ratio of the likelihood evaluated with background only to the likelihood evaluated including the source of interest. Using this test statistic as the detection criterion, the LCR estimates the observed LAT flux for sources with TS $\geq$ 4  and uses a Bayesian profile likelihood method to calculate upper limits for intervals yielding TS $<$ 4. There are multiple ways to download the LCR data for offline analysis. The light curve data for individual sources can be downloaded in CSV or JSON formats through their specific Source Report pages.

\section{Results}
\subsection{Temporal Analysis}
\begin{figure*}[t]
    \centering
    \includegraphics[angle=0, width=0.4\textwidth]{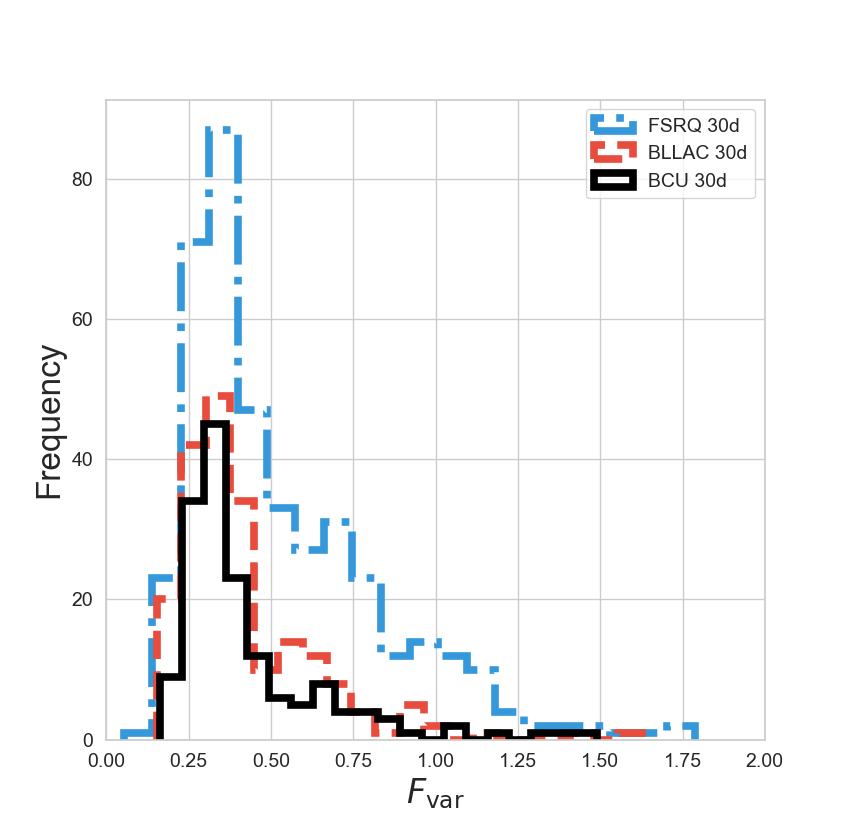}\hspace{-0.1em}
    \includegraphics[angle=0, width=0.4\textwidth]{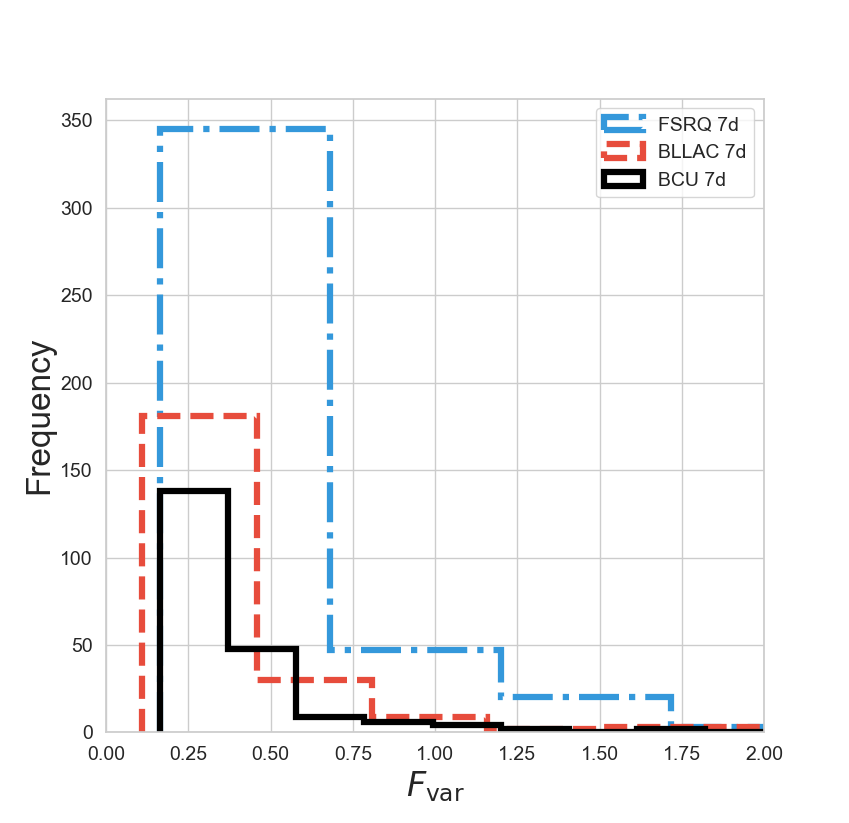}\hspace{-0.1em}
    \includegraphics[angle=0, width=0.4\textwidth]{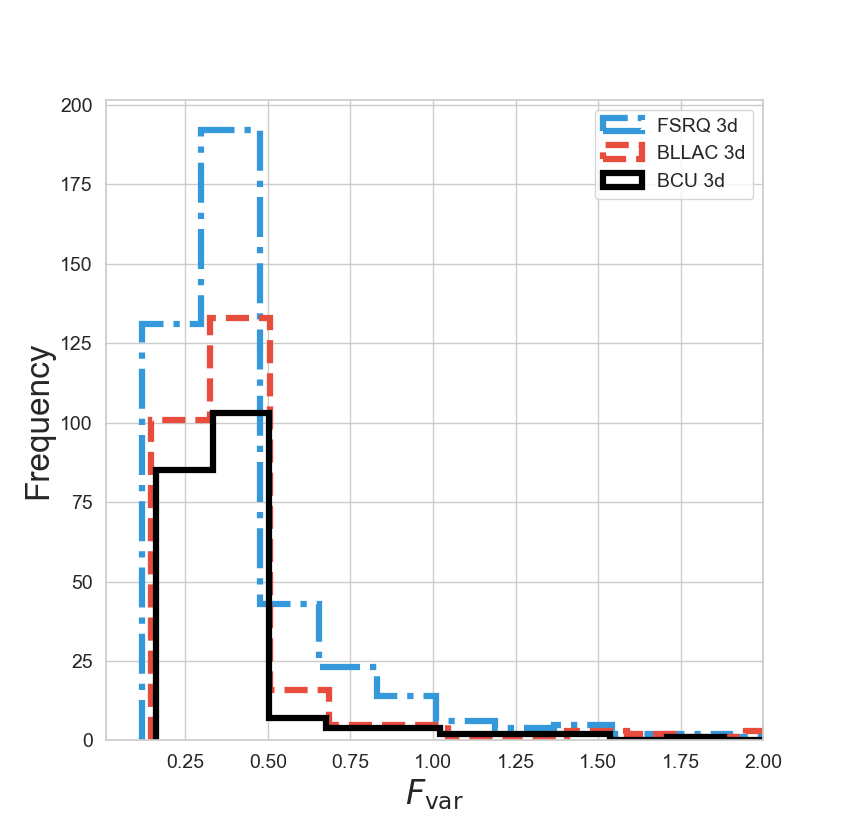}
    \caption{ Distribution of $F_{\text{var}}$ for FSRQs, BL\,Lacs, and BCUs. The three panels represent histograms of light curves binned over 3-day, 7-day, and 30-day intervals. The black solid histogram corresponds to BCUs, the red dashed histogram to BL Lacs, and the blue dash-dotted histogram to FSRQs. The results indicate that FSRQs generally exhibit higher variability compared to BL Lacs and BCUs.}
    \label{Fig:fvar}
\end{figure*}

In this section, we present a detailed temporal analysis of the 3-day, 7-day and 30-day bin light curves of blazars acquired from the {\emph Fermi}-LAT LCR,  alongside flux values, we have also used the index values from the LCR. These light curves encompass an energy range of 0.1--100 GeV.  We have focused on data points where TS $>$ 4 ($2\sigma$ detection) and $flux/flux_{err}>2$ to ensure the maximum data retention.  After applying these criteria, the number of sources that remain in the (3-day, 7-day, and 30-day) bins are as follows: FSRQs (573, 572, 565), BL Lacs (456, 437, 389), and BCUs (335, 334, 302). The analysis in different time bins will help in understanding the effects of binning on the variability of the source, since almost every light curve we observe has some sort of binning associated. The large sample of light curves of blazars available in the three time bins offers a good platform for this kind of study.
The variability amplitude is  important parameter that can be deduced from the light curves, which in turn can provide constraints on the physical processes that cause $\gamma$-ray flux variations. Therefore, we calculated the fractional variability amplitude ($\rm F_{var}$) as defined in \cite{2003MNRAS.345.1271V} using the equation:
\begin{equation}\label{eq:fvar}
F_{\text{var}} = \sqrt{\frac{S^2 - \overline{\sigma_{\text{err}}^2}}{\overline{F}^2}},
\end{equation}

where $S^2$ represents the variance of the fluxes, $\overline{F}$ is the mean flux, and $\overline{\sigma_{\text{err}}^2}$ is the  mean square error of the fluxes. The uncertainty on $F_{\text{var}}$ is determined using the equation:
\begin{equation}
F_{\text{var,err}} = \sqrt{\frac{1}{2N}\left(\frac{\overline{\sigma_{\text{err}}^2}}{F_{\text{var}}\overline{F}^2}\right)^2 + \frac{1}{N}\frac{\overline{\sigma_{\text{err}}^2}}{\overline{F}^2}},
\end{equation}

where $N$ is the number of points in the light curve. In Figure \ref{Fig:fvar}, we show the  histograms of the fractional variability for FSRQs, BL Lacs and BCUs objects in the three time bins (3-day, 7-day, and 30-day). In the histograms, we have considered the blazars for which the $F_{var}/F_{var,error}>2$.  The mean fractional variability values for FSRQs, BL\,Lac and BCU objects are given in Table \ref{tab:fvar_corr}. Though there is no clear distinction between the histograms of FSRQs, BL\,Lacs and BCUs in all the time bins, however, the mean values of $\rm F_{var}$ reveal that FSRQ objects tend to exhibit higher variability compared to BL\,Lac across all three time bins. In 3-day binned $\gamma$-ray lightcurves,  BCUs mean variability value  are closer to those of FSRQ.  However, in the 7-day and 30-day binned $\gamma$-ray light curves, the BCU mean variability  are closer to the  BL\,Lacs.
These results imply that there are effects  of binning on the $\rm F_{var}$ values. 
 Further, to statistically test the differences, a two-sample Kolmogorov-Smirnov (KS) test was applied to the $F_{var}$ values at the 95\% confidence level. The comparison of the KS statistic between FSRQs, BL\,Lacs, and BCUs across different time bins is presented in Table \ref{tab:fvar_corr}. 
These results demonstrate a statistically significant difference in variability distributions between FSRQs and BL\,Lacs, as well as between FSRQs and BCUs, particularly in the 7-day and 30-day time bins. However, the null hypothesis probability values rules out the differences between BL\,Lacs and BCUs.  The KS test results between BL\,Lacs and BCUs suggest no significant variability differences, indicating that BL\,Lacs and BCUs  may exhibit similar variability patterns in the observed time bins. This suggests that fractional variability is useful for distinguishing FSRQs from BL Lacs, but less effective in differentiating BL Lacs from BCUs. 
The higher variability observed in FSRQs could be linked to intrinsic physical processes, such as more active central engines or distinct jet structures compared to BL Lacs. 

\begin{table*}[t]
\caption{The top table presents the weighted mean ${\rm F_{var}}$ values for FSRQs, BL\,Lacs and BCUs, derived from light curves binned at 3-day, 7-day, and 30-day intervals, with selection criteria of  TS $>$ 4 and $Flux/Flux_{err}>2$.  The bottom table presents the KS test p-values used to assess the statistical significance of differences in distributions between the FSRQs, BL\,Lacs and BCUs for the light curves, based on the same binning intervals and selection thresholds as in top table.}

\centering
\begin{tabular}{|c|c|c|c|c|}
\hline
 & \multicolumn{4}{c|}{Fvar} \\
\cline{2-5}
& Blazar Type & 3-day  & 7-day  & 30-day  \\
\cline{3-5}
\multirow{2}{*}{Mean $F_{var}$} & FSRQs    & 0.96 $\pm$ 0.001 & 1.02 $\pm$ 0.001 & 1.07 $\pm$ 0.001 \\
& BL Lacs  & 0.64 $\pm$ 0.002 & 0.66 $\pm$ 0.002 & 0.62 $\pm$ 0.002 \\
& BCUs     & 0.80 $\pm$ 0.004 & 0.81 $\pm$ 0.004 & 0.81 $\pm$ 0.004 \\
\hline
 & \multicolumn{4}{c}{KS-statistic (p-value)} \\
\cline{2-5}
& & 3-day Bin & 7-day Bin & 30-day Bin \\
\cline{3-5}
\multirow{2}{*}{KS-statistic } & FSRQs vs BL Lacs & 0.11 (p = 0.03) & 0.16 (p = 4.3e-4) & 0.20 (p = 1.85e-5) \\
comparison & FSRQs vs BCUs    & 0.12 (p = 0.01) & 0.21 (p = 3.10e-6) & 0.24 (p = 9.94e-6) \\
& BL Lacs vs BCUs  & 0.08 (p = 0.30) & 0.09 (p = 0.19)    & 0.08 (p = 0.64) \\
\hline
\end{tabular}
\label{tab:fvar_corr}
\end{table*}

\subsection{Effect of time-binning on the $F_{var}$}
To further investigate the effect of binning on $F_{\mathrm{var}}$ values,  we analyzed the $F_{\mathrm{var}}$ values for FSRQs, BL Lacs, and BCUs as a function of time binning intervals--specifically, 3-day, 7-day, and 30-day bins. Only statistically significant flux points were included in the $F_{\mathrm{var}}$ calculation to ensure accurate results. 
 In Table \ref{tab:fvar_corr}, we have shown the weighted average values of the $F_{var}$ for FSRQs, BL\,Lacs, and BCUs.  The average \( F_{var} \) values clearly illustrate distinct variability behaviors among the blazar classes as a function of the time-binning interval; FSRQs exhibit a steady increase in \( F_{var} \) from \( 0.96\pm0.001 \) at the 3-day bin to \( 1.07\pm0.001 \) at the 30-day bin, indicating that their variability amplitude is more pronounced on longer timescales, whereas BL\,Lacs show a modest rise from \( 0.64\pm0.002 \) to \( 0.66\pm0.002 \) between the 3-day and 7-day bins, followed by a decrease to \( 0.62\pm0.002 \) at the 30-day bin, suggesting a dominance of shorter-term variability; BCUs, on the other hand, maintain a relatively constant \( F_{var} \) (around \( 0.80\pm0.004 \) to \( 0.81\pm0.004 \)) across all binning. The FSRQ variability, pattern across time bins is consistent with previous study  such as \citet{2024ApJ...977..111A}, the authors reported  that $F_{\mathrm{var}}$ of FSRQ  source PKS\,0805-07 tends to increase with larger bin sizes when data points with $TS > 4$ are considered. In contrast, \citet{2019Galax...7...62S} work on BL\,Lac source showed that the increase in bin size decreases the $F_{\mathrm{var}}$. However, the authors considered all data points  regardless of uncertainties. FSRQs are comparatively brighter than BL\,Lacs at $\gamma-ray$ energy. To investigate the role of source brightness, we selected the three bright and three faint FSRQs and BL\,Lacd based on the mean flux values from the 30-day binned light curves. Table~\ref{tab:fsrq_fvar_bright} and \ref{tab:blac_fvar_bright} presents the mean $F_{\mathrm{var}}$ values for these selected bright and faint FSRQs and BL,Lac sources. 
 For bright FSRQs, $F_{\mathrm{var}}$ exhibits an increasing trend with larger bin sizes, rising from 3-day to 30-day intervals. While for bright BL\,Lac sources, $F_{\mathrm{var}}$ increases from 3-day to 7-day bins but then declines as the bin size extends to 30 days. These results are consistent with the average behavior of these two classes. Additionally, the errors on $F_{\mathrm{var}}$ in bright sources remain small and largely independent of bin size. In contrast, for the faint FSRQs, the trend in $F_{\mathrm{var}}$ is less consistent and fluctuate without a clear trend. This is due to the reduction in the larger number of  data points from the light curves and therefore is  a direct manifestation of photon‐statistics bias.  In either cases the $F_{var}$ values changing with the time suggests that we should maintain consistent time binning. This is critical when comparing variability across different energy bands to ensure reliable conclusions. In summary, these trends underscore two points:photon‐statistics bias strongly affects $F_{\rm var}$ in faint sources, and consistent time‐binning is critical when comparing variability across energy bands, to avoid misinterpreting bin‐size effects as intrinsic physical differences.

\begin{table*}[t]
\caption{ Summary of $F_{\mathrm{var}}$ values for the three brightest and three faintest FSRQs across 3-day, 7-day, and 30-day binned light curves. The light curves contains time bins having TS $>$ 4 and $Flux/Flux_{err} > 2$. Each time-bin column includes sub-columns for the number of flux points, mean flux, and $F_{\mathrm{var}}$ values.}
\centering
\begin{tabular}{|c|c|ccc|ccc|ccc|}
\hline
\textbf{FSRQ Type} & \textbf{Source} & \multicolumn{3}{c|}{\textbf{3-day}} & \multicolumn{3}{c|}{\textbf{7-day}} & \multicolumn{3}{c|}{\textbf{30-day}} \\
\cline{3-11}
 &  & Number & mean flux & $F_{var}$ & Number & mean flux & $F_{var}$ & Number & mean flux & $F_{var}$ \\
\hline
\multirow{3}{*}{Bright FSRQ} 
& 4FGL\,1427.9-4206  & 1491 & 6.57e-7 & $0.78\pm 0.004$  & 706 &  6.08e-7 &  $0.82\pm 0.005$ & 180 & 5.65e-7 & $0.82\pm 0.004$ \\
& 4FGL\,1833.6-2103  & 1403 & 1.14e-6 & $1.45 \pm 0.007$ & 724 & 9.77e-7 & $1.50\pm 0.006$ & 185 & 9.27e-7 & $1.50\pm 0.005$ \\
& 4FGL\,2253.9+1609  & 1633 & 1.40e-6 & $1.42\pm 0.002$   & 752 & 1.30e-6 & $1.43\pm 0.002$ & 184 & 1.33e-6 & $1.60\pm 0.002$ \\
\hline
\multirow{3}{*}{Faint FSRQ} 
& 4FGL\,2023.6-1139   & 10 & 2.43e-7 & $0.41\pm 0.18$ & 9 & 1.33e-7 & $0.33\pm 0.25$  &  7 & 4.46e-8 & $0.51\pm 0.28$  \\
& 4FGL 2050.4-2627  & 17 & 1.63e-7 & $0.39\pm 0.16$  & 12 &  1.14e-7 & $0.73\pm 0.15$  &  10 & 3.85e-8 & $0.30 \pm 0.17$\\
& 4FGL\,1615.6+4712   & 8 & 1.00e-7 & $0.38\pm 0.22$ & 8 & 6.84e-8 & $0.43\pm 0.20$ &      5 &  2.01e-8 & $0.27\pm 0.33$   \\
\hline
\end{tabular}
\label{tab:fsrq_fvar_bright}
\end{table*}

\begin{table*}[t]
\caption{Summary of $F_{\mathrm{var}}$ values for the three brightest and three faintest FSRQs across 3-day, 7-day, and 30-day binned light curves. The light curves contains time bins having TS $>$ 4 and $Flux/Flux_{err} > 2$. Each time-bin column includes sub-columns for the number of flux points, mean flux, and $F_{\mathrm{var}}$ values.}
\centering
\begin{tabular}{|c|c|ccc|ccc|ccc|}
\hline
\textbf{BL\,Lac Type} & \textbf{Source} & \multicolumn{3}{c|}{\textbf{3-day}} & \multicolumn{3}{c|}{\textbf{7-day}} & \multicolumn{3}{c|}{\textbf{30-day}} \\
\cline{3-11}
 &  & Number & mean flux & $F_{var}$ & Number & mean flux & $F_{var}$ & Number & mean flux & $F_{var}$ \\
\hline
\multirow{3}{*}{Bright BL\,Lac} 
& 4FGL\,2202.7+4216  & 1530 & 5.32e-7 & $0.93\pm 0.004$  & 742 &  4.84e-7 &  $0.95\pm 0.004$ & 187 & 4.55e-7 & $0.90\pm 0.004$ \\
& 4FGL\,0721.9+7120 & 1498 & 2.17e-7 & $0.55 \pm 0.006$ & 754 & 1.94e-7 & $0.61\pm 0.006$ & 187 & 1.86e-7 & $0.56\pm 0.006$ \\
& 4FGL\,1104.4+3812  & 1713 & 1.89e-7 & $0.38\pm 0.007$   & 779 & 1.82e-7 & $0.40\pm 0.007$ & 187 & 1.79e-7 & $0.37\pm 0.006$ \\
\hline
\multirow{3}{*}{Faint BL\,Lac} 
& 4FGL\,0051.2-6242   & 11 & 7.11e-8 & $0.24\pm 0.29$ & 13 & 3.43e-8 & $0.18\pm 0.34$  &  38 & 1.13e-8 & $0.24\pm 0.12$  \\
& 4FGL\,1150.6+4154  & 14 & 1.09e-7 & $0.09\pm 0.53$  & 14 &  3.79e-8 & $0.46\pm 0.18$  &  47 & 1.06e-8e-8 & $0.18 \pm 0.14$\\
& 4FGL\,1610.7-664   & 11 & 2.87e-7 & $0.99\pm 0.19$ & 8 & 1.23e-7 & $0.88\pm 0.29$ &      38 &  1.11e-8 & $0.21\pm 0.16$   \\
\hline
\end{tabular}
\label{tab:blac_fvar_bright}
\end{table*}

\section{Flux-index corelation}
\begin{figure*}[t]
    \centering
    \includegraphics[angle=0, width=0.4\textwidth]{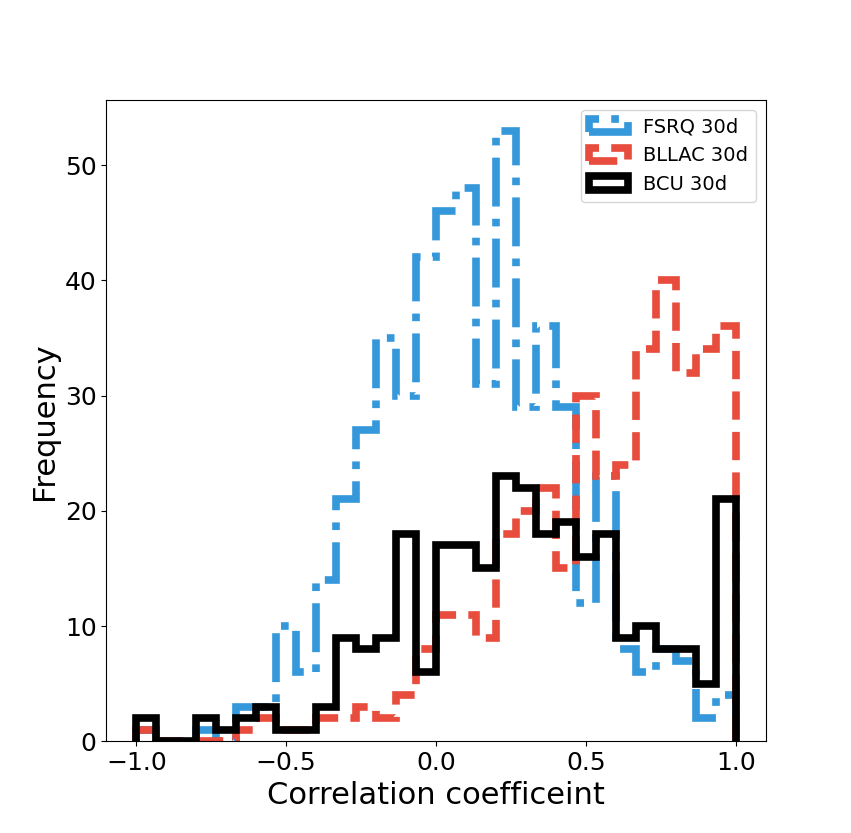}\hspace{-0.1em}
    \includegraphics[angle=0, width=0.4\textwidth]{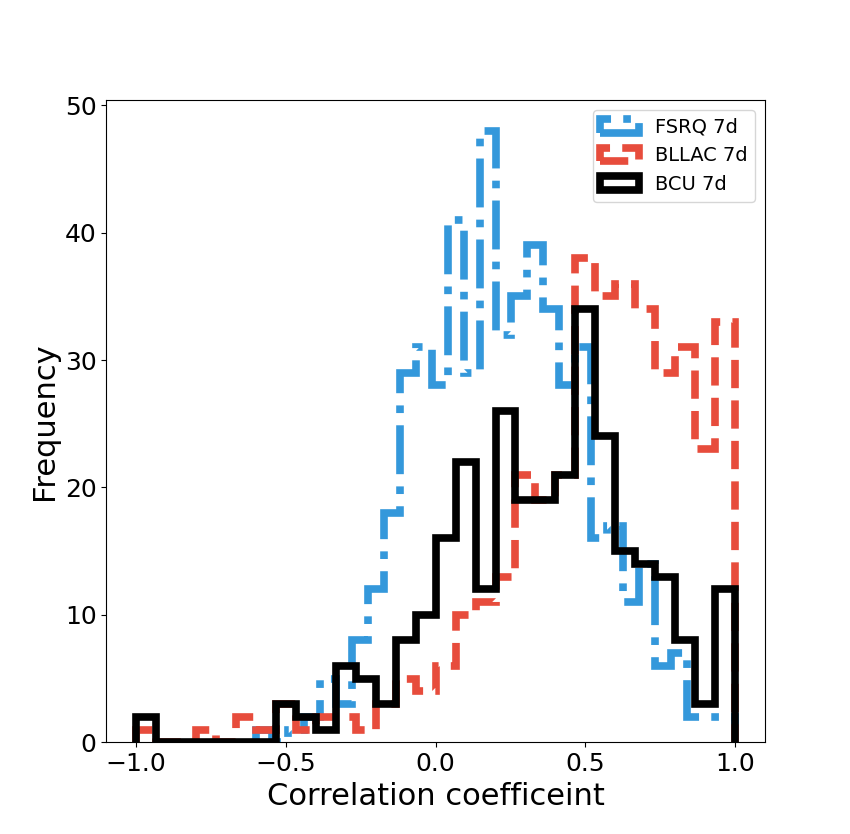}\hspace{-0.1em}
    \includegraphics[angle=0, width=0.4\textwidth]{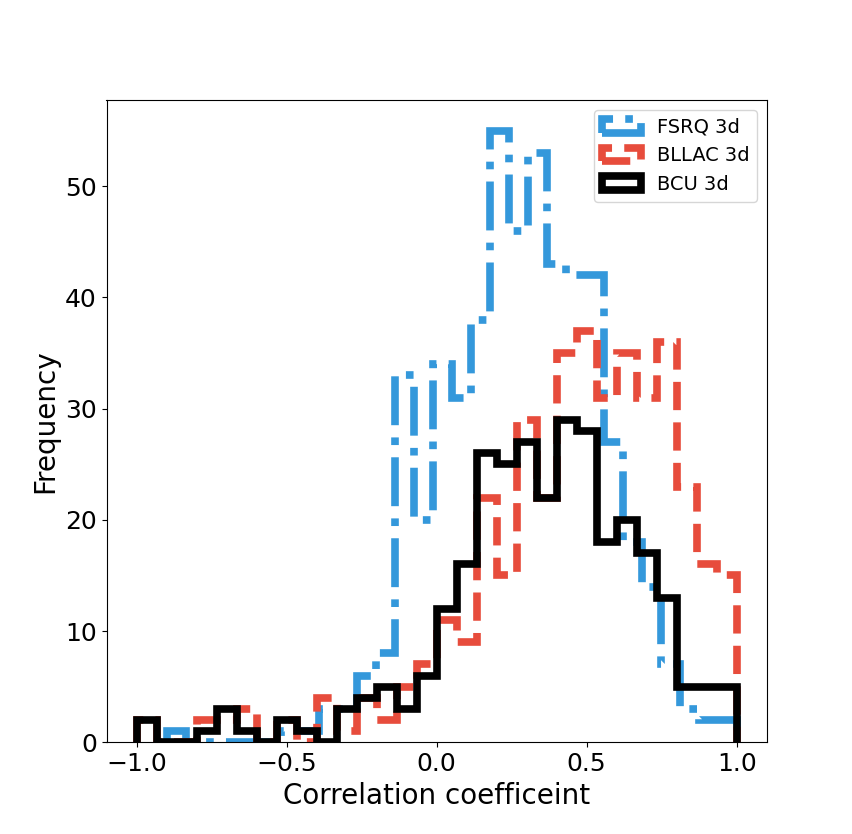}
    \caption{ Distribution of the correlation coefficient between flux and spectral index for FSRQs, BL Lacs, and BCUs using 30-day (top left), 7-day (top right), and 3-day (bottom) binned light curves. Only time bins with TS$>$ 4, $Flux/Flux_{err} > 2$, and $Index/Index_{err} > 2$ are included. The results show a predominantly positive correlation for BL Lacs and BCUs, while FSRQs tend to exhibit negative correlations across all time bins.}
    \label{fig:rank_corr_coef}
\end{figure*}

\begin{figure*}[t]
    \centering
    \includegraphics[angle=0, width=0.3\textwidth]{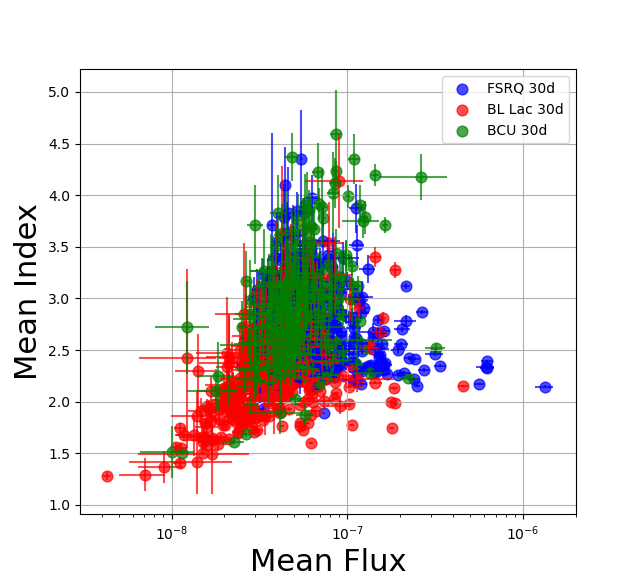}\hspace{-0.1em}
    \includegraphics[angle=0, width=0.296\textwidth]{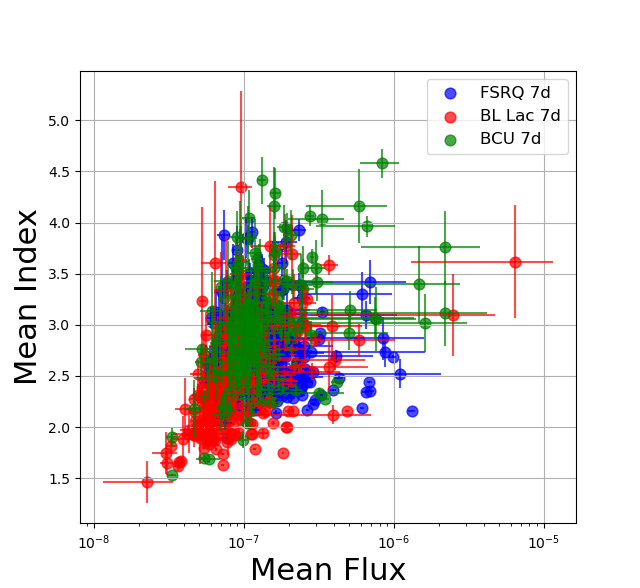}\hspace{-0.1em}
    \includegraphics[angle=0, width=0.3\textwidth]{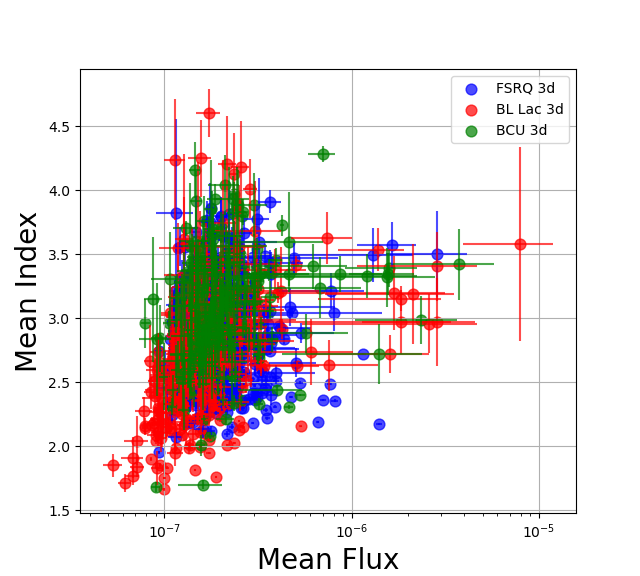}
    \caption{ Scatter plots showing the relationship between mean spectral index and mean flux for three time-binned light curves: 30-day (left), 7-day (middle), and 3-day (right), for FSRQs, BL Lacs, and BCUs. The results suggest a mild positive correlation between index and flux for  BL Lacs, and BCUs, whereas FSRQs exhibit an anti-correlation trend. }
    \label{fig:scatter_mean_index_flux}
\end{figure*}

The {\emph Fermi} LCR provides the  index information apart from the flux values discussed in previous section. We considered the index and flux values where the ratio of index to its error and the ratio of flux to its error were greater than 2.  We also excluded time periods with an index value exceeding 6 to remove outliers points.  We used the Spearman rank correlation method to investigate the expected correlation between index and flux values for all the sources in the 3 days, 7 days and 30 days binned light curves. This method gives us a rank coefficient and a probability value for the null hypothesis. In the Figure \ref{fig:rank_corr_coef}, we displayed  the distribution of  rank coefficients for the three types of blazars across three time periods. The results indicate that the spectral index and flux are positively correlated for most BL Lac and BCU sources across all time bins. Similarly, FSRQs exhibit a negative correlation between the index and flux, albeit with a reduced correlation magnitude. To explore this further, we calculated the mean flux and mean index for each source. Figure \ref{fig:scatter_mean_index_flux} displays scatter plots of the mean index versus mean flux for FSRQ, BL Lac, and BCU sources across the time bins.
The Spearman rank correlation analysis between the mean flux and mean spectral index for  BL Lacs, FSRQs, and BCUs, across 3-day, 7-day, and 30-day time bins are presented in Table {\ref{tab:spe}}  \footnote{ While the Spearman rank correlation coefficient is below 0.5 in most cases, the corresponding low p-values indicate that the trends, albeit weak, are statistically significant and not likely due to random fluctuations. This is a consequence of the large sample size, which increases the sensitivity of statistical tests}.
These values reveal distinct trends among FSRQS, BL Lacs, and BCUs. BL\,Lacs show a moderate positive correlation in all bins, with $r_s \approx 0.46-0.47$ and  P-values ($P_s \approx 10^{-22}-10^{-23}$), indicating that as the flux increases, these sources become spectrally softer. If the spectral index  becomes more negative, then the high-energy photons are fewer relative to lower energies.  
This can be explored under the physical scenario, where the acceleration process is more efficient, but the cooling is also stronger, leading to a balance where the electron distribution has a steeper slope.
BCUs exhibit a similar mild positive correlation, with $r_s$ values ranging from 0.25 - 0.36 and P-values ($P_s \approx 10^{-6} - 10^{-10}$), implying a comparable trend of spectral softening with increasing flux. In contrast, FSRQs show a slight anticorrelation trend, with $r_s$ values between -0.02 and -0.12, suggesting a marginal tendency for these sources to become spectrally harder as flux increases. While P-values for the 7-day and 30-day bins reject the nul-hypothesis that the index-flux are not correlated, for the 3-day bin lightcurve the nul-hypothesis can not be rejected.   The lack of correlation at the 3-day bin indicates that short-term variability may obscure any consistent flux-index relationship in FSRQs. These results indicate that blazars exhibit complex and varied behavior depending on the type of blazar and the timescale of observation.
This could reflect different physical mechanisms active over longer timescales, with changes in particle acceleration, cooling, or the surrounding environment influencing spectral properties. The cumulative behavior of blazars, integrating both flaring and quiescent states over extended periods, underscores the importance of multi-timescale observations to fully understand the variability and emission processes in these sources.


\begin{table*}[t]
\centering
 \caption{Spearman rank correlation results ($r_s$ and $P_s$) between mean flux and mean spectral index for BL Lacs, FSRQs, and BCUs across 3-day, 7-day, and 30-day binned light curves. Column 1 lists the blazar type; Columns 2, 3, and 4 show results for 3-day, 7-day, and 30-day bins, respectively, with each containing the correlation coefficient ($r_s$) and p-value ($P_s$) as sub-columns.}
\begin{tabular}{|c|cc|cc|cc|}
\hline
\cline{1-7}
Blazar Type & \multicolumn{2}{c|}{3-day Bin} & \multicolumn{2}{c|}{7-day Bin} & \multicolumn{2}{c}{30-day Bin} \\
\cline{2-7}

& $r_s$ & $P_s$ & $r_s$ & $P_s$ & $r_s$ & $P_s$ \\
\hline

BL Lacs & 0.46 & $3.98 \times 10^{-22}$ & 0.47 & $8.87 \times 10^{-23}$ & 0.47 & $4.33 \times 10^{-23}$ \\
FSRQs   & -0.02 & $0.69$ & -0.11 & $8.30 \times 10^{-3}$ & -0.12 & $4.85 \times 10^{-3}$ \\
BCUs    & 0.25 & $9.08 \times 10^{-6}$ & 0.34 & $1.53 \times 10^{-9}$ & 0.36 & $3.87 \times 10^{-10}$ \\
\hline

\end{tabular}
\label{tab:spe}
\end{table*}

\subsection{Flux Distribution}
The examination of long-term flux distributions in the light curves of astrophysical systems is a valuable approach for understanding the physical processes driving variability. A normal flux distribution typically points to additive processes, whereas a lognormal distribution suggests the influence of multiplicative processes. In compact black hole systems, the flux distribution is generally observed to follow a lognormal pattern. To investigate this behavior in blazars, we analyzed the $\gamma$-ray flux distribution across three time bins (3-day, 7-day, and 30-day) using the Anderson-Darling (AD) test and histogram fitting. The AD test provides a test statistic (TS) value, and if the TS exceeds the critical value (CV) at a 5\% significance level, the normality of the flux distribution is rejected.

\begin{table*}[t]
\caption{ Number of sources showing log-normality and the number of sources where log-normality was rejected for FSRQs, BL\,Lacs and BCUs across 3-day, 7-day, and 30-day binned light curves. Column 1 lists
the blazar type; Columns 2 and 3 are meant for sources in which lognormality is accepted and rejected respective, with each containing subcolumns representng number of sources in the 3 day, 7 day and 30 day. }
\centering
\begin{tabular}{|c|c|c|c|c|c|c|}
\hline
\multirow{2}{*}{Blazar Type} & \multicolumn{3}{c|}{LN accepted} & \multicolumn{3}{c|}{LN rejected} \\
\cline{2-7}
 & $N_{3}$ & $N_{7}$ & $N_{30}$ & $N_{3}$ & $N_{7}$ & $N_{30}$ \\
\hline
BL Lacs & 326 & 326 & 335 & 130 & 111 & 54 \\
FSRQs  & 311 & 314 & 395 & 262 & 258 & 170 \\
BCUs    & 234 & 249 & 243 & 101 & 85  & 59 \\
\hline
\end{tabular}
\label{tab:log_normality_sources}
\end{table*}

We considered the flux points in the light curve where the fitting was convergent and included the points for which TS was greater than 4. Additionally, the ${\rm flux/flux_{error} > 2}$ was also included in the analysis to remove the flux points with large errors. Moreover, we removed the outlier points by choosing the flux values with values less than $10^{-9}$.  We used the Anderson-Darling (AD) test to confirm the log-normality of flux distributions in the blazar lightcurves. The number of sources showing log-normality is mentioned in the Table \ref{tab:log_normality_sources}. For the sources which showed log-normal distribution, we fitted the normalized histograms in log-scale with a Gaussian distribution, described by the equation

\begin{equation}\label{eq:ln}
L(x) = \frac{1}{\sqrt{2\pi}\sigma} e^{-(x-\mu)^2/2\sigma^2}
\end{equation}

\begin{figure*}[t]
    \centering
    \includegraphics[angle=0, width=0.3\textwidth]{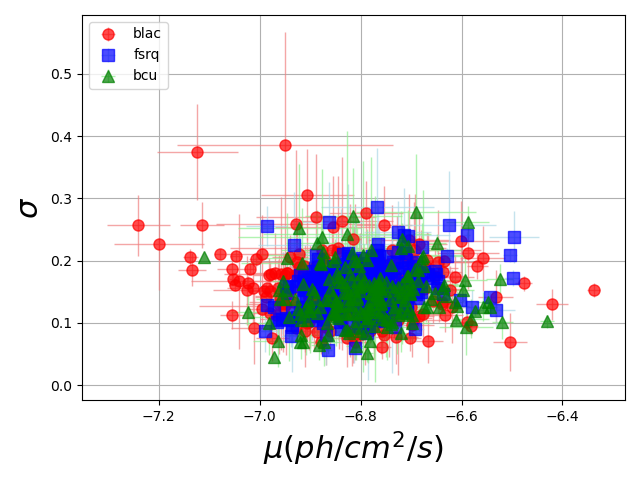}\hspace{-0.1em}
    \includegraphics[angle=0, width=0.3\textwidth]{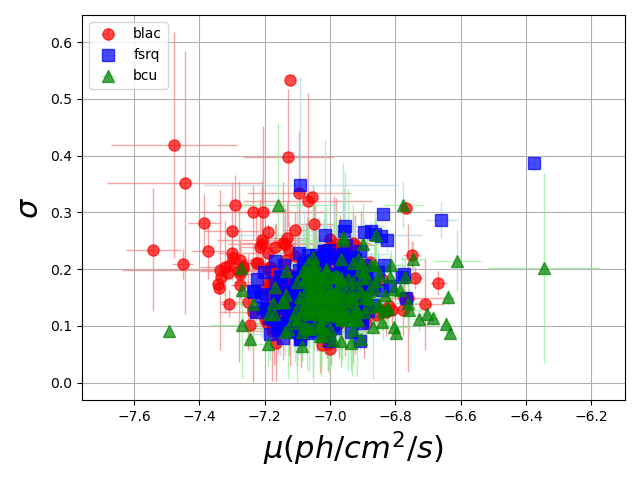}\hspace{-0.1em}
    \includegraphics[angle=0, width=0.3\textwidth]{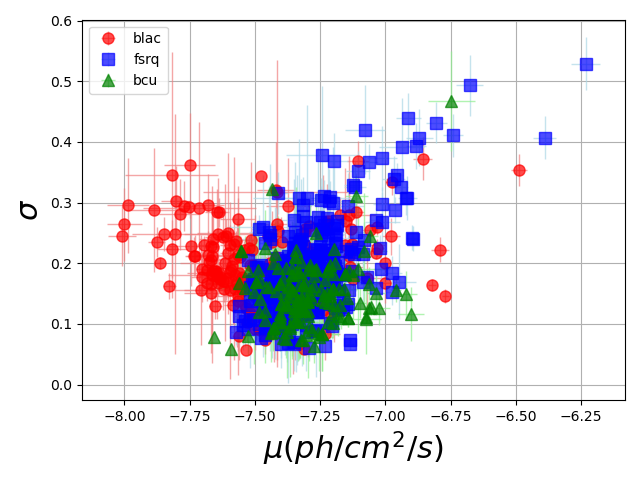}
    \caption{ Scatter plots of the log-normal fit parameters $\mu$ and $\sigma$ for FSRQs, BL,Lacs, and BCUs, based on flux distributions from 30-day (left), 7-day (middle), and 3-day (right) binned light curves. BL\,Lac sources exhibit greater scatter compared to FSRQs and BCUs.}
    \label{fig:scat_mu_sigma_lognor}
\end{figure*}

This equation fitted to normalized histogram in log-scale results in lognormal fit. Here $\mu$ and $\sigma$ are the centroid and width of the logarithm of flux  distribution.  We analysed the scatter plot between $\mu$ and $\sigma$, obtained from the lognormal fit (see Figure \ref{fig:scat_mu_sigma_lognor}). In the 3-day binned light curves, no clear distinction is observed among BL\,Lac, FSRQ, and BCU sources. However, BL,Lac sources exhibit a broader scatter compared to FSRQs and BCUs, which is consistent with the fact that BL\,Lacs encompass three subclasses: LBLs, IBLs, and HBLs.   The Spearman rank correlation results ($r_s$ and $P_s$ values)  across the three binned light curves are given in Table \ref{tab:mu_sigma_lognor}. 
The results suggest that for BL\,Lacs, there is no correlation between $\mu$ and $\sigma$ at the 3-day bin, but a moderate positive correlation emerges at the 7-day bin, indicating some relationship between the parameters on this timescale, though it weakens again at the 30-day bin. In contrast, FSRQs show a progressively stronger positive correlation as the time bin increases, suggesting a greater relationship between $\mu$ and $\sigma$ over longer timescales.
BCUs, however, exhibit no significant correlation across any time bin, indicating that the centroid and width of the lognormal fit remain largely independent for this class of blazars. 
This variability behavior highlights distinct differences in how $\mu$ and $\sigma$ relate across different blazar classes and timescales.

\begin{table*}[t]
\caption{Spearman rank correlation results ($r_s$ and $P_s$ values) between  $\mu$ and $\sigma$ values obtained  from the log-normal fit for BL Lacs, FSRQs, and BCUs, based on 3-day, 7-day, and 30-day binned light curves.}
\centering
\begin{tabular}{c|cc|cc|cc}
\hline
\multicolumn{7}{c}{Correlation between $\mu$ and $\sigma$ (lognormal fit)} \\
\cline{1-7}
Blazar Type & \multicolumn{2}{c|}{3-day Bin} & \multicolumn{2}{c|}{7-day Bin} & \multicolumn{2}{c}{30-day Bin} \\
\cline{2-7}

& $r_s$ & $P_s$ & $r_s$ & $P_s$ & $r_s$ & $P_s$ \\
\hline

BL Lacs & -0.06 & 0.29 & 0.23 & $2 \times 10^{-4}$ & 0.11 & 0.08 \\
FSRQs  & 0.21  & $2 \times 10^{-4}$ & 0.34 & $2 \times 10^{-9}$ & 0.45 & $2 \times 10^{-18}$ \\
BCUs    & 0.09  & 0.19 & 0.09 & 0.20 & 0.10 & 0.21 \\
\hline
\end{tabular}
\label{tab:mu_sigma_lognor}
\end{table*}

Out of the total sources considered in our analysis, the number of sources where log-normality is rejected is shown in Table \ref{tab:log_normality_sources}. We checked the flux distribution of these sources  with the bi-model PDF defined by

\begin{equation}\label{eq:nor}
D(x) = \frac{a}{\sqrt{2\pi}\sigma_1} e^{-(x-\mu_1)^2/2\sigma_1^2}+ \frac{1-a}{\sqrt{2\pi}\sigma_2} e^{-(x-\mu_2)^2/2\sigma_2^2}
\end{equation}

 where a  is the mixing fraction that determines the relative contributions of the two components. We further examined the correlations among the fitting parameters of the double lognormal distributions. The Spearman rank correlation analysis between the double lognormal parameters ($\mu_1$, $\mu_2$, $\sigma_1$, $\sigma_2$) for BL\,Lacs, FSRQs, and BCUs--evaluated across 3-day, 7-day, and 30-day light curves--is summarized in Table \ref{tab:double_log_correlation} and reveals the following relationships.   For FSRQs, $\mu_1$ and $\mu_2$ consistently show mild positive correlations across all time bins, with the maximum correlation at 7 days ($r_s = 0.41$, $P_s = 4.37 \times 10^{-6}$), and $\mu_1$ vs. $\sigma_1$ as well as $\mu_2$ vs. $\sigma_2$ also exhibit  positive correlations, particularly at shorter time scales. In BL\,Lacs, $\mu_1$ and $\mu_2$ exhibit a significant positive correlation in the 3-day light curve ($r_s = 0.57$, $P_s = 2.6 \times 10^{-4}$), which weakens at longer time scales, while $\sigma_1$ and $\sigma_2$ show a significant negative correlation at 30 days ($r_s = -0.64$, $P_s = 4.9 \times 10^{-3}$).
 In BCUs, $\mu_1$ and $\mu_2$ show significant positively correlation at 3 days ($r_s = 0.65$, $P_s = 7.8 \times 10^{-4}$), aligning with the trend observed in BL\,Lacs. However, $\mu_2$ and $\sigma_2$ have a significant negative correlation at 3 days ($r_s = -0.55$, $P_s = 5.9 \times 10^{-3}$). 
 These correlations suggest varying relationships between the lognormal parameters across different blazar classes and time scales, potentially reflecting different underlying physical mechanisms working at different time scales.

\begin{table*}[t]
\centering
\caption{ Spearman rank correlation results ($r_s$ and $P_s$ values) between the best fit parameters $\mu_1$, $\mu_2$, $\sigma_1$ and $\sigma2$ obtained  from the double log-normal fit for BL Lacs, FSRQs, and BCUs, based on 3-day, 7-day, and 30-day binned light curves.}
\label{tab:double_log_correlation}
\begin{tabular}{lccccc}
\toprule
Blazar type & parametrs & \multicolumn{3}{c}{$r-value$ and $p-value$} \\
\cmidrule(lr){3-5}
 & & 3 Day & 7 Day & 30 Day \\
\midrule
BL\,Lac & $\mu_1$ vs. $\mu_2$ & $0.57$, $2.6\times10^{-4}$ & $0.27$, $0.18$ & $0.18$, $0.48$ \\
& $\sigma_1$ vs. $\sigma_2$ & $-7.7\times10^{-4}$, $0.99$ & $-0.36$, $0.07$ & $-0.64$, $4.9\times 10^{-3}$ \\
& $\mu_1$ vs. $\sigma_1$  & $-0.03$, $0.83$ & $-0.06$, $0.77$ &  $-0.04$, $0.85$\\
& $\mu_2$ vs. $\sigma_2$  & $0.25$, $0.13$ & $0.29$, $0.15$ & $0.05$, $0.82$ \\
\midrule
FSRQ & $\mu_1$ vs. $\mu_2$ & $0.36$, $5.5\times10^{-5}$ & $0.41$,$4.37\times10^{-6}$ & $0.27$, $0.02$ \\
& $\sigma_1$ vs. $\sigma_2$ & $-0.18$, $0.05$ & $-0.01$, $0.96$ & -0.33, $3.2\times 10^{-3}$ \\
& $\mu_1$ vs. $\sigma_1$ & $0.48$, $2.59\times10^{-8}$ & $0.50$,  $8\times 10^{-9}$ & $0.43$, $1\times 10^{-4}$\\
& $\mu_2$ vs. $\sigma_2$ & $0.48$, $3.43\times10^{-8}$ & $0.43$, $1.11\times 10^{-6}$ & $0.03$, $0.76$\\
\midrule
BCU & $\mu_1$ vs. $\mu_2$ & $0.65$, $7.8\times10^{-4}$ & $0.47$, $0.02$ & $0.58$, $0.04$ \\
& $\sigma_1$ vs. $\sigma_2$ & $-0.02$, $0.92$ & $-0.09$, $0.70$& $-0.35$, $0.23$ \\
& $\mu_1$ vs. $\sigma_1$ & $0.34$, $0.11$ & $0.43$, $0.04$ & $-0.28$, $0.35$\\
& $\mu_2$ vs. $\sigma_2$ &  $-0.55$, $5.9\times 10^{-3}$ & $-0.04$, $0.86$ & $0.04$, $0.88$\\
\bottomrule
\end{tabular}
\end{table*}

\subsection{Effect of time binning on the flux distribution}
The availability of light curves in the three time bins facilitated an examination of the effect of binning on the observed flux distributions. We analyzed the log-normality behavior of sources across 3-day, 7-day, and 30-day binned light curves for BL Lacs, FSRQs, and BCUs. The analysis focuses on the  number of sources exhibiting log-normality, including sources that are lognormal to a particular bin light curve or lognormal across multiple binned light curves. Our results show significant overlap in the log-normality distributions between different time binned light curves, although some sources demonstrate unique behavior depending on the binning interval. Table \ref{tab:total_common_unique_sources} provides a comprehensive summary of the total number of sources in each bin, along with the common and unique sources observed between pairs of time bins. The log-normality of blazar sources across different time bins reveals that while a significant number of sources exhibit consistent log-normality in multiple time bins, the 30-day bin tends to introduce more unique sources, especially in FSRQs. This suggests that longer time-scale observations capture different variability patterns than shorter ones, which could be attributed to different underlying physical processes. nterestingly, \citet{2018RAA....18..141S} demonstrated that when a light curve contains a sufficiently large number of data points, binning does not significantly affect the log-normal nature of the flux distribution—provided the power spectral density (PSD) has a non-zero spectral slope. To test this, we examined the number of sources exhibiting lognormality in light curves binned over 3, 7, or 30 days. Our focus was particularly on sources that are uniquely lognormal in one specific binning scheme and contain more than 300 significant data points. This analysis serves to verify the consistency between simulated light curve results reported by \citet{2018RAA....18..141S} and our observational findings. Further details are provided in the next section.

\begin{table*}[t]
\centering
\caption{Total sources, common, and unique sources showing log-normality across different binned light curves for BL Lacs, FSRQs, and BCUs. The first column shows the blazar type, second column shows the total number of sources in each time bin (3-day, 7-day, 30-day) for BL Lacs, FSRQs, and BCUs. The rest of the columns show the common and unique sources between each combination of time bins (3-day \& 7-day, 3-day \& 30-day, 7-day \& 30-day). Unique sources are listed in parentheses, showing how many are unique to each bin.}
\scalebox{0.9}{ 
\begin{tabular}{c|c|c|c|c|c|c|c}
\hline
Blazar Type & Total (3-day, 7-day, 30-day) & \multicolumn{2}{c|}{3-day \& 7-day} & \multicolumn{2}{c|}{3-day \& 30-day} & \multicolumn{2}{c}{7-day \& 30-day} \\
\cline{3-8}
 & & Common & Unique (3-day, 7-day) & Common & Unique (3-day, 30-day) & Common & Unique (7-day, 30-day) \\
\hline
BL Lacs & 326, 326, 330 & 244 & (82, 80) & 238 & (88, 97) & 250 & (74, 85) \\
FSRQs   & 311, 314, 395 & 230 & (81, 84) & 235 & (76, 160) & 250 & (64, 145) \\
BCUs    & 234, 249, 243 & 192 & (42, 57) & 173 & (61, 70) & 192 & (57, 51) \\
\hline
\end{tabular}
}
\label{tab:total_common_unique_sources}
\end{table*}


\section{PSD}
We analyzed the PSD of the $\gamma$-ray light curve, which exhibits log-normal flux distributions. To enhance statistical reliability, we focused on sources with more than 300 data points after implementing the quality cuts outlined earlier. 
More data points allowed the PSD to capture accurate details in the frequency domain.
After applying the cuts, we identified 14 and 10 sources in BL Lacs and FSRQs  respectively  possessing lognormality in the 3-day binned $\gamma$-light curve, and 18 and 8 sources in BL Lacs and FSRQs respectively  possessing lognormality in the 7-day binned $\gamma$-light curve, respectively. However, no BCU sources met the threshold, and no BL Lac or FSRQ sources qualified for the 30-day binned light curve.
 The PSDs are obtained using the Bartlett’s method, splitting the light curve into equal-length segments, calculating the periodogram in each, and then averaging them into the final periodogram. We fitted the power  law model to all averaged PSDs, the higher frequency were limited to 0.01 1/day,  the fit parameters for the PSDs of the sources are summarized in Table \ref{psd_blac_fsrq}. 
 In all the PSDs, we noted no evidence  of a break. The average PL slope obatined in BL\,Lac and FSRQs are obtained  as $0.43\pm 0.07$ and $0.54\pm 0.10$, respectively in 3 days binned light curves, while $0.52 \pm 0.09$ and $0.85 \pm 0.09$ in 7 days binned light curves. A power law PSD suggests that  there is no single characteristic timescale that dominates the behavior—instead, fluctuations occur over a wide, continuous range of timescales.  This is a common signature of turbulent processes and shock-induced emission. The rebinning of lightcurves from 3-day to 7-day bins, effectively average out  the short-timescale (high-frequency) variations. As a result, the relative power from longer timescales (low frequencies) becomes more prominent. A steeper power-law index  indicates that low-frequency (longer timescale) fluctuations contribute more to the total variance than high-frequency ones.
 
For BL\,Lac sources showing lognormal behavior, we observed that the PSD index is less than 1 in 3 day. Notably, three sources --4FGL\,0449.4-4350, 4FGL\,1104.4+3812, and 4FGL\,1653.8+3945 -- are log-normal in the 3-day binned light curve but do not show single lognormal behavior in the 7-day binned light curve. The PSD of these three sources shows spectral slopes of $0.71 \pm 0.08$, $0.69 \pm 0.07$, and $0.26 \pm 0.09$, respectively. The 7-day binned light curve rejects the log-normality of these three sources and  instead their flux distributions are better fitted with a double lognormal PDF, yielding reduced $\chi^2$ values of 0.47, 1.05, and 0.50. These results indicate that log-normality  though bimodel is still consistent within these sources. 
In the 7-day binned light curve, seven sources namely 4FGL\,1555.7+1111, 4FGL\,1427.0+2348, 4FGL\,1217.9+3007, 4FGL\,1015.0+4926, 4FGL\,0538.8-4405, 4FGL\,0211.2+1051, and 4FGL\,1248.3+5820 exhibit lognormal behavior, which is not observed in the 3-day binned light curves. The PSD slopes of these sources have a wide spread from 0.15--1.40 with 4FGL\,0538.8-4405 showing a steep slope, which has PSD slopes of $1.40 \pm 0.15$. The flux distribution in the 3-day binned light curves for these sources are better fitted by a double lognormal PDF. 

Among the 11 BL Lacs showing log-normality in both the 3-day and 7-day light curves, we found that the PSD slope of 8 BL Lacs is less than 0.5, with a few consistent with white noise light curves. For these sources, the transition from 3-day to 7-day binning does not alter the log-normality. This result is consistent with \citet{2020MNRAS.496.3348S}, which showed that the effect of binning becomes significant when averaging the data $\sim 8$ times the baseline time interval, allowing log-normality to be rejected at the 10\% significance level, and completely rejected if we average the data 64 times the baseline time interval. 
For FSRQs, among the 10 sources showing log-normality in the 3-day binned light curves, only 4FGL 2236.3+2828 remains lognormal in the 7-day binned light curve. The sources that are not log-normal in the 7-day binned light curve exhibit double log-normality. Similarly four FSRQs, which show log-normality in  7 day binned light curve, but not in the  3 day binned light curve, their distribution in the three day binned light curve is double lognormal. 
It should be noted that  there can be  number of effects that can, potentially, distort the PDS of our analysis from the “true” long-term variability pattern. This includes stochastic variability within a finite length of observation, systematic in the data, and statistical noise.

\begin{table*}[t]
\centering
\caption{ PSD analysis results of the 3-day and 7-day binned $\gamma$-ray light curves, each with more than 300 data points and showing log-normal flux distributions in either the 3-day binning, the 7-day binning, or both. The upper half of the table lists results for BL,Lac sources, while the lower half corresponds to FSRQ sources. The left-side columns present the results for the 3-day binning, and the right-side columns for the 7-day binning. Columns: (1) Source name; (2,3,4) normalization, slope, and reduced $\chi^2$ from the power-law (PL) fit for the 3-day binning light curves; (5,6,7) corresponding parameters for the 7-day binning.}
\begin{adjustbox}{max width=\textwidth}
\small
\begin{tabular}{lccc|ccc}
\toprule
\multirow{2}{*}{Source} & \multicolumn{3}{c|}{BL\,Lac (3 day bin)} & \multicolumn{3}{c}{BL\,Lac (7 day bin)} \\
\cmidrule(lr){2-4} \cmidrule(lr){5-7}
 & Norm & alpha & $\chi^2/\text{dof}$ & Norm & alpha & $\chi^2/\text{dof}$ \\
\midrule
4FGL\,1517.7-2422 & $1.94\pm 0.29$ & $0.53\pm 0.10$ &  1.14 & $4.02\pm 0.49$ & $0.40\pm 0.19$ & 1.82 \\
4FGL\,1543.0+6130 & $2.95\pm 0.55$ & $0.15\pm 0.13$ & 2.27 & $2.20\pm 0.38$ & $0.14\pm 0.26$ & 1.06 \\
4FGL\,0222.6+4302 & $3.95\pm 0.66$ & $0.72\pm 0.11$ & 2.1 & $4.34\pm 0.45$ & $0.91\pm 0.13$ & 3.22 \\
4FGL\,0112.1+2245  & $1.61\pm 0.27$ & $0.44\pm 0.12$ & 1.94 & $5.09\pm 0.60$ & $0.93\pm 0.19$ & 1.69 \\
4FGL\,0449.4-4350 &  $3.31\pm 0.39$ & $0.71\pm 0.08$ &  1.29 & & & \\
4FGL\,1104.4+3812 & $1.71\pm 0.18$ & $0.69\pm 0.07$ & 1.22 & & & \\
4FGL\,2000.0+6508 &  $1.21\pm 0.25$ & $0.49\pm 0.14$ & 1.23 &  $2.07\pm 0.29$ & $0.32\pm 0.25$  & 1.60 \\
4FGL\,1058.4+0133 & $2.50\pm 0.72$ & $0.69\pm 0.20$ & 1.41 & $5.11\pm 0.80$ & $0.80\pm 0.25$  & 2.60 \\
4FGL\,2139.4-4235 & $2.28\pm 0.34$ & $0.27\pm 0.10$ & 0.84 & $3.08\pm 0.57$ & $0.32\pm 0.28$  &  1.19 \\
4FGL\,1653.8+3945 & $0.81\pm 0.12$ & $0.26\pm 0.09$ & 1.44 & & & \\
4FGL\,2158.8-3013 &  $3.31\pm 0.37$ & $0.83\pm 0.07$ & 0.93 & $4.83 \pm 0.45$ & $0.98\pm 0.12$  & 1.27\\
4FGL\,0144.6+2705 & $1.25\pm 0.18$ & $0.10\pm 0.09$ & 1.31 & $2.16 \pm  0.31$ & $0.02\pm 0.22$ & 6.47 \\
4FGL\,0818.2+4222 & $1.88\pm 0.28$ & $0.11 \pm 0.13$ & 0.65 & $1.76 \pm 0.23$ & $0.04\pm 0.17$ & 0.46 \\
4FGL\,1806.8+6949  &  $1.13\pm 0.21$ & $0.10\pm 0.13 $ & 0.83 &  $1.81 \pm 0.27$ & $0.32\pm 0.21$ & 0.89 \\
4FGL\,1555.7+1111 & & & & $2.92\pm 0.28$ & $0.62\pm 0.15$ & 1.52 \\
4FGL\,1427.0+2348 & & & & $2.54\pm 0.27$ & $0.22\pm 0.16$ & 0.97 \\
4FGL\,1217.9+3007 & & & & $3.46\pm 0.41$ & $0.15\pm 0.18$ & 1.11 \\
4FGL\,1015.0+4926 & & & & $2.82\pm 0.32$ & $0.52\pm 0.20$ & 2.00 \\
4FGL\,0538.8-4405 & & & & $3.22\pm 0.32$ & $1.40\pm 0.15$  &  3.72\\
4FGL\,0211.2+1051 &  & & & $3.02\pm 0.48$ & $0.43 \pm 0.26 $ & 0.57 \\
4FGL\,1248.3+5820 & & & & $3.47\pm 0.54$ & $0.86\pm 0.26 $  & 4.71 \\
\midrule
 & \multicolumn{3}{c|}{FSRQ (3 day bin)} & \multicolumn{3}{c}{FSRQ (7 day bin)} \\
\cmidrule(lr){2-4} \cmidrule(lr){5-7}

4FGL\,0532.6+0732 & $4.73\pm 0.75$ & $0.84\pm 0.11$ & 0.43 & & & \\
4FGL\,1127.0-1857 & $2.06\pm 0.50$ & $0.67\pm 0.16$ & 1.39 & & & \\
4FGL\,0526.2-4830 & $1.94\pm 0.44$ & $0.54\pm 0.15$ & 1.47 & & & \\
4FGL\,0217.8+0144 & $1.40\pm 0.23$ & $0.06\pm 0.12$ & 0.86 & & & \\
4FGL\,0457.0-2324 & $3.04\pm 0.33$ & $0.93\pm 0.07$ & 1.23 & & & \\
4FGL\,1246.7-2548 & $3.93\pm 0.59$ & $0.53\pm 0.09$ & 0.87 & & & \\
4FGL\,0453.1-2806 & $2.33\pm 0.40$ & $0.10\pm 0.11$ & 1.04 & & & \\
4FGL\,2236.3+2828 & $1.94\pm 0.45$ & $0.14\pm 0.15$ & 1.80 & $5.85\pm 1.03$ & $0.95\pm 0.29$ & 1.27 \\
4FGL\,2345.2-1555 & $6.28\pm 1.53$ & $1.01\pm 0.15$ & 1.29 & & & \\
4FGL\,0407.0-3826 & $2.68\pm 0.43$ & $0.58 \pm 0.11$ & 2.21 & & & \\

4FGL\,2348.0-1630 & & & & $10.38\pm 1.59$ & $0.90\pm 0.15$ & 1.06 \\
4FGL\,2229.7-0832 & & & & $6.70\pm 1.25$ & $0.70\pm 0.18$ & 1.84\\
4FGL\,2356.4+4030 & & & & $5.83\pm 0.83$ & $0.56\pm 0.11$ & 0.57 \\
4FGL\,1057.8+0138 & & & & $4.99\pm 1.02$ & $0.65\pm 0.14$ & 0.63 \\
4FGL\,0909.1+0121 & & & & $6.62\pm 1.38$ & $0.77\pm 0.16$ & 0.80 \\
4FGL\,1345.5+4453 & & & & $8.16\pm 0.88$ & $1.45\pm 0.10$ & 1.79\\
4FGL\,1256.1-0547 &  &  &  &  $20.02\pm 1.86$  & $0.80\pm 0.08$ & 3.76 \\

\bottomrule
\end{tabular}
\end{adjustbox}
\label{psd_blac_fsrq}
\end{table*}

\section{Spectral Index Distribution}
 Given the narrow spread of the photon index distribution for each blazar type, significant changes in the photon index over time are not expected for a given source. Even though flux variations are substantial (by a factor of more than 7) in blazars, the photon index range is only about $\sim 0.3$ wide, accounting for a $1\sigma$ dispersion. 
We analyzed the spectral index distribution for blazars exhibiting lognormal flux distributions, selecting sources with more than 100 data points in their light curves. 
 Importantly, we found that the corresponding index distributions deviate from a Gaussian distribution, instead exhibiting either log-normal or double log-normal behavior. Our results indicate that  the index values of FSRQ  sources mostly follow a double log-normal distribution (see Table \ref{table:index_dist}). Specifically, in the 3-day binning interval, 125 FSRQ sources exhibit a log-normal index distribution, while 147 show a double log-normal distribution. This trend is less pronounced in the 7-day binned light curves, where 72 sources show a log-normal index distribution and 104 show a double log-normal distribution. However, BL\,Lac and BCU sources show a mixed trend in the three time bins (3-day, 7-day, and 30-day).
 
\begin{figure*}[t]
    \centering
    \includegraphics[width=0.32\linewidth]{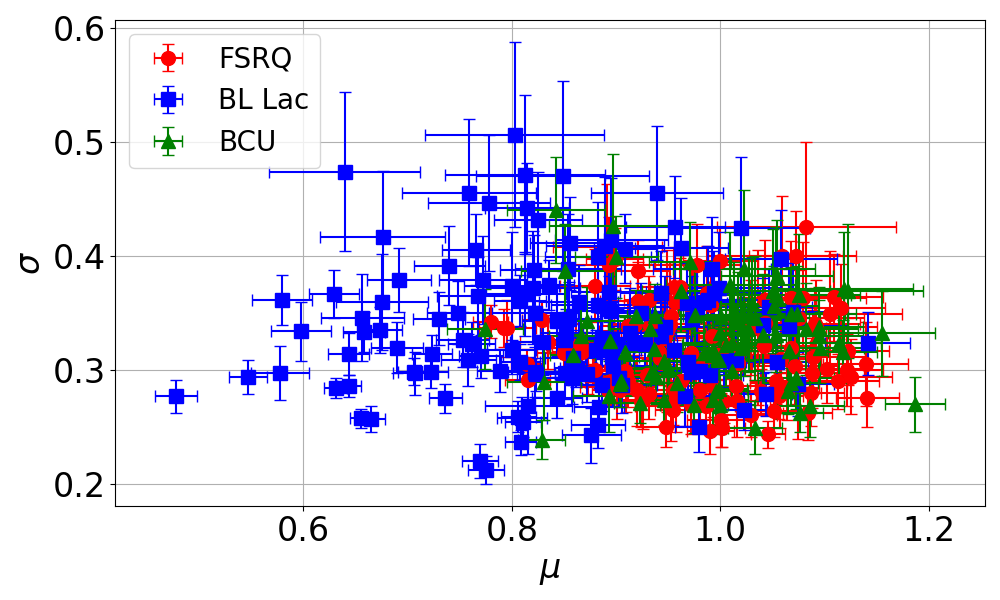}
    \hfill
    \includegraphics[width=0.32\linewidth]{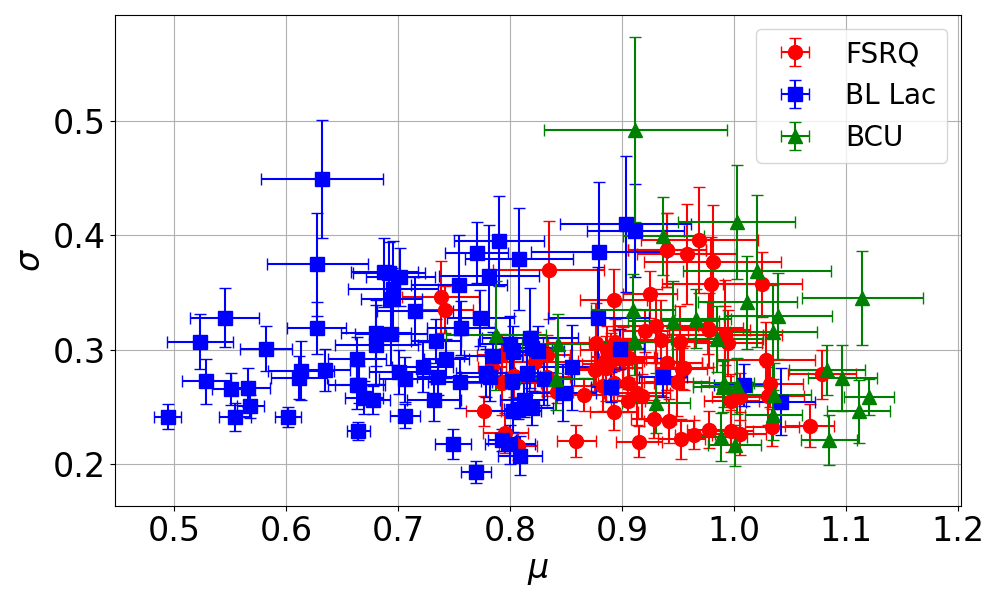}
   \hfill   \includegraphics[width=0.32\linewidth]{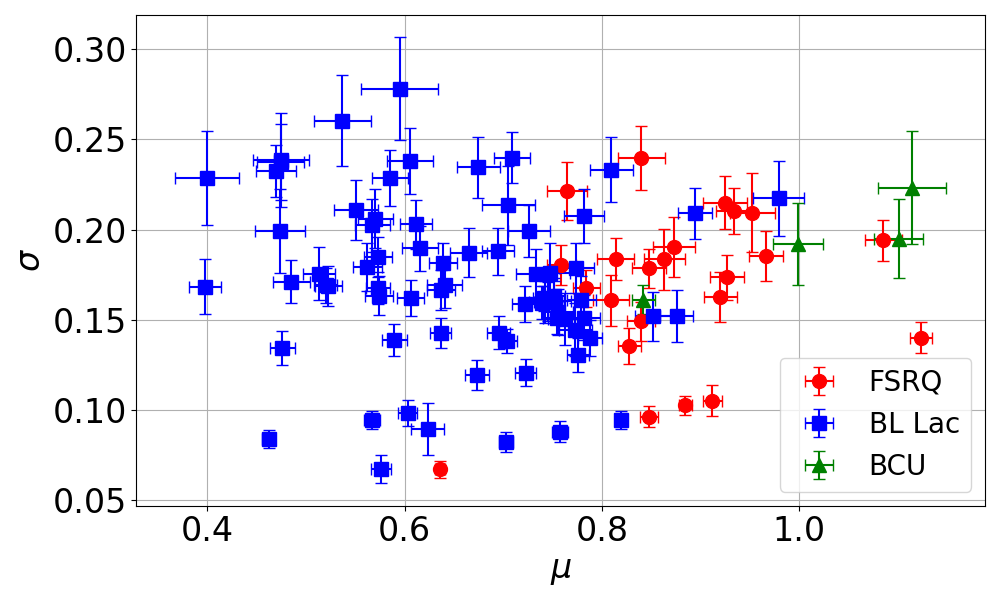}
      \caption{The scatter plot shows the best-fit values of $\mu$ and $\sigma$ obtained from lognormal fits to the spectral index distributions for the three blazar classes (BL Lacs, BCUs, and FSRQs) across three time binnings: 3-day (left), 7-day (middle), and 30-day (right).}
   \label{fig:index_dist}
   \end{figure*}
\begin{figure*}[t]
    \centering
    \includegraphics[width=0.32\linewidth]{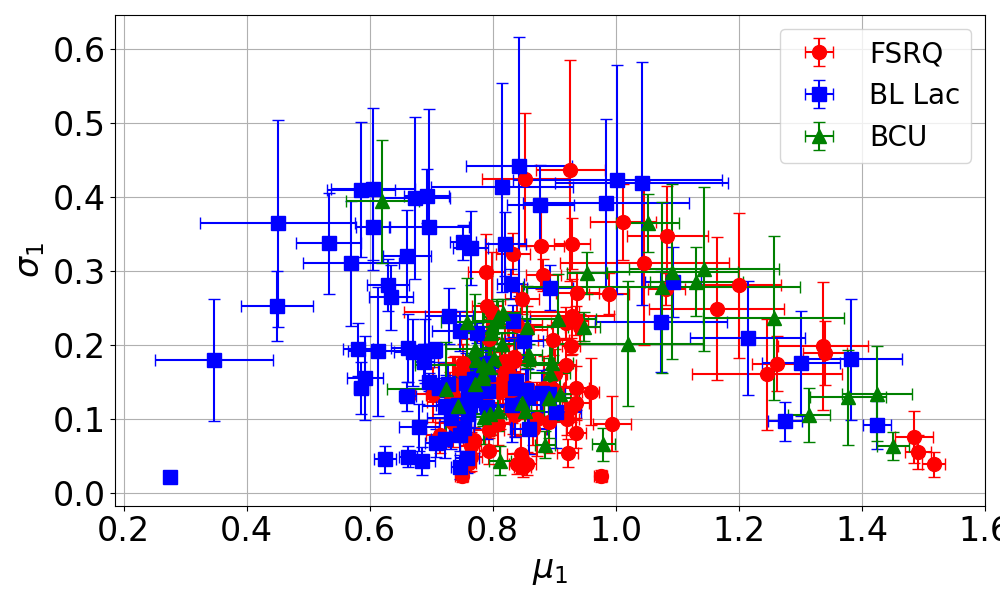}
    \hfill
    \includegraphics[width=0.32\linewidth]{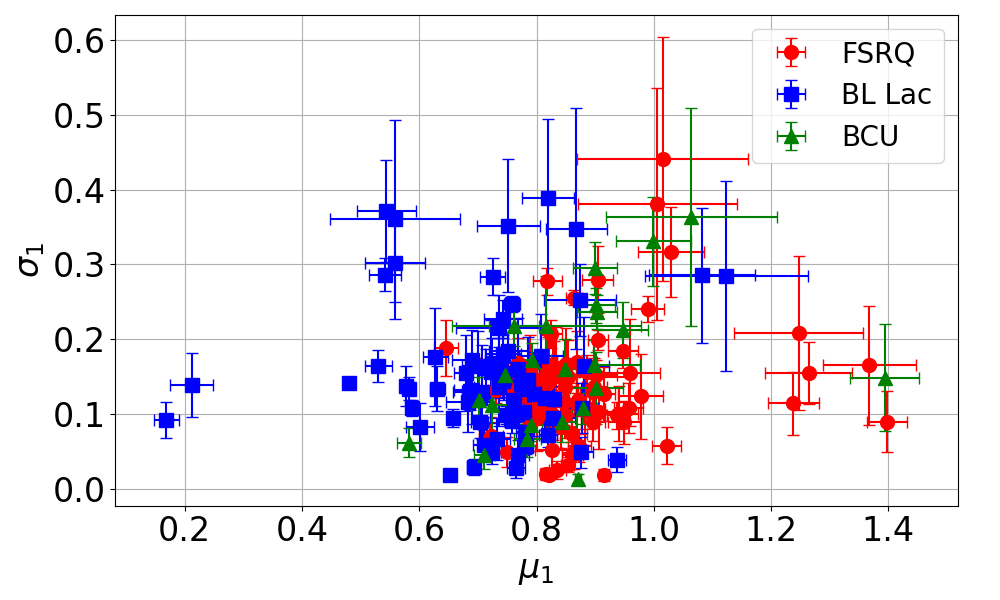}
    \hfill
    \includegraphics[width=0.32\linewidth]{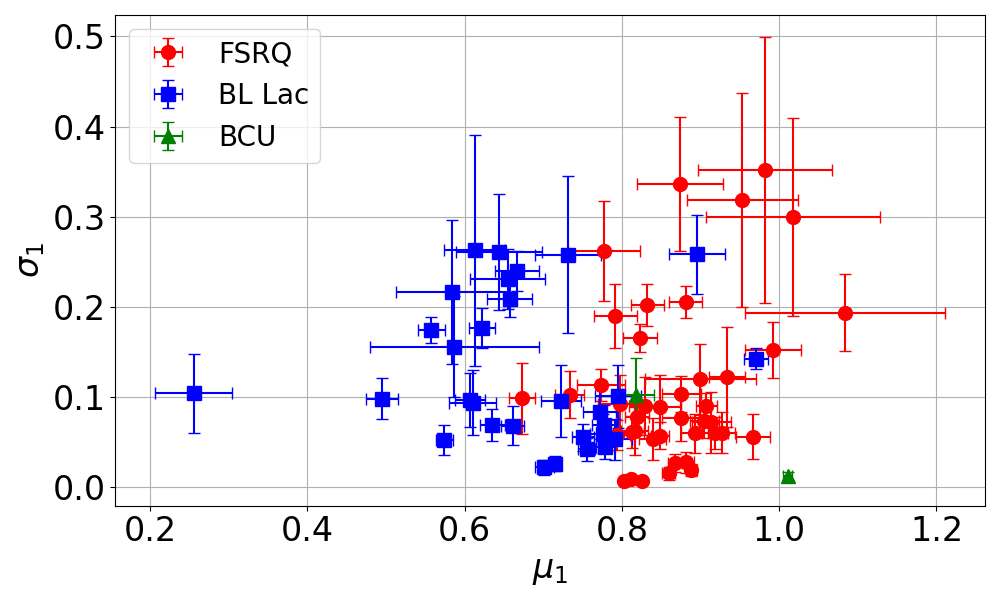}
    \hfill
    \includegraphics[width=0.32\linewidth]{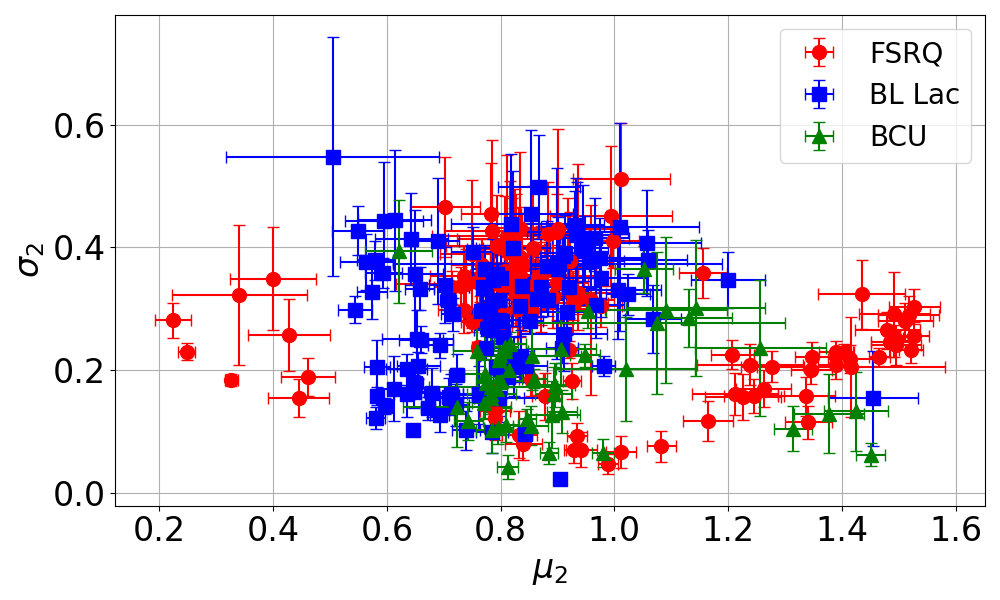}
    \hfill
    \includegraphics[width=0.32\linewidth]{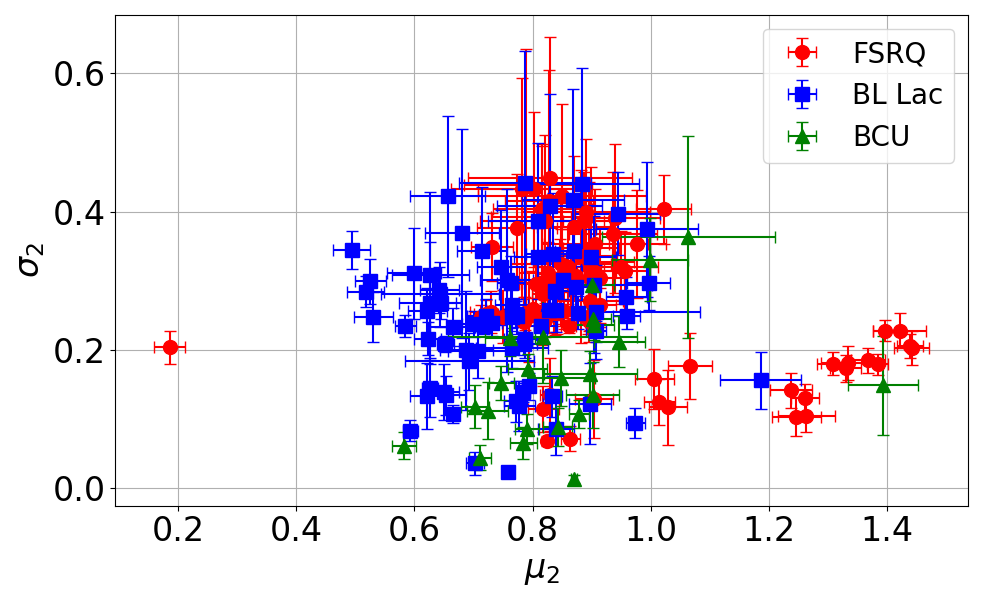}
    \hfill
    \includegraphics[width=0.32\linewidth]{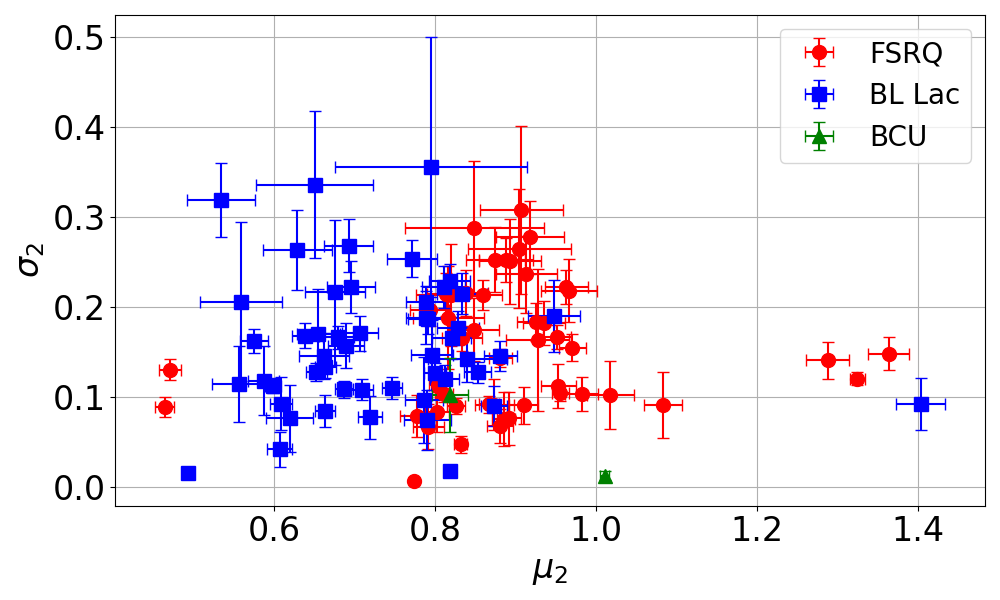}
    
  \caption{ The scatter plot between the $\mu_1$ and $\sigma_1$ values (top panel), and $\mu_2$ and $\sigma_2$ values (bottom panel) derived from the double log-normal fit for the index distributions of FSRQs, BL Lacs, and BCUs across the three time bins:  3-day (left), 7-day (middle), and 30-day (right).}
    \label{fig:index_dist_double}
    \end{figure*}

In addition to this, we also plotted the scatter between the $\mu$ and $\sigma$ values obtained from the log-normal fit to the index distribution for the FSRQs, BL\,Lacs and BCUs across the three time bins (3-day, 7-day, and 30-day) in the Figure \ref{fig:index_dist}.  The scatter plot in case of BL\,Lacs shows broader range values of centriod and sigma values compared to the FSRQs and BCUs. This is expected as BL\,Lac sources are composed of HBL, IBL and LBL sources. 
 This observation aligns with our  findings from the flux distribution. Additionally, we examined the scatter between the $\mu_1$ and $\sigma_1$ values, as well as the $\mu_2$ and $\sigma_2$ values derived from the double log-normal fit for the index distributions of FSRQs, BL Lacs, and BCUs across the three time bins (see Figure \ref{fig:index_dist_double}). While the plots show a slight difference in scatter between FSRQs and BL Lacs, there is no definitive evidence indicating whether BCUs are more aligned with FSRQs or BL Lacs. 

 These findings suggest that the variability in spectral index distributions is not random but instead follows distinct patterns in FSRQs, BL\,Lacs and BCUs. The dominance of double log-normal distributions, especially in FSRQs, points to a more complex variability mechanism at $\gamma$-ray energies compared to the simpler log-normal distributions observed in flux. This result diverges from previous studies of blazar behavior at X-ray energies \citep{2020MNRAS.491.1934K}, where the index distribution was found to be Gaussian while the flux distribution followed a log-normal pattern. The Gaussian index distribution at X-ray energies implies that perturbations in the acceleration timescale follow a Gaussian process, which in turn leads to the observed log-normal flux distribution. In contrast, the observation of log-normal index distributions at $\gamma$-ray energies suggests that flux variations in these sources are not directly tied to variations in the spectral index, indicating potentially different emission mechanisms or variability processes at higher energies.

\begin{table*} 
\caption{Summary of sources showing log-normal and double log-normal index distributions across different binning intervals.}
\centering
\begin{tabular}{|c|c|c|c|c|c|}
\hline
\textbf{Source Type} & \textbf{Binning Interval} & \textbf{Log-normal (No. of Sources)} & \textbf{Double Log-normal (No. of Sources)} \\ \hline
\multirow{3}{*}{FSRQ}   & 3-day  & 125  & 147  \\ \cline{2-4} 
                        & 7-day  & 72   & 104  \\ \cline{2-4} 
                        & 30-day & 23   & 67   \\ \hline
\multirow{3}{*}{BL Lac} & 3-day  & 135  & 123  \\ \cline{2-4} 
                        & 7-day  & 78   & 101  \\ \cline{2-4} 
                        & 30-day & 72   & 66   \\ \hline
\multirow{3}{*}{BCU}    & 3-day  & 82   & 75   \\ \cline{2-4} 
                        & 7-day  & 19   & 32   \\ \cline{2-4} 
                        & 30-day &  4    & 4    \\ \hline
\end{tabular}
\label{table:index_dist}
\end{table*}

\section{Summary and Discussion}
\label{sum:discussion}

In this study, we carried a systematic analysis of long-term $\gamma$-ray data from the {\emph Fermi}-LAT for blazars  comprising  FSRQs, BL Lacs and BCUs  to characterize their $\gamma$-ray variability. We focussed on 3-day, 7-day, and 30-day binned light curves to assess how binning impacts variability. Using data where TS $>$ 4 (a $2\sigma$ detection threshold), we examined $\rm F_{var}$ by constucting histograms of  $\rm F_{var}$ across time bins. The results show that FSRQs exhibit higher mean variability compared to BL Lacs and BCUs, while BCUs demonstrate intermediate variability. A KS test reveals differences between FSRQs and BL Lacs, and between FSRQs and BCUs, particularly in 7-day and 30-day bins, while differences between BL Lacs and BCUs are not significant. These findings suggest that $\rm F_{var}$ may distinguish FSRQs from BL Lacs, but less so for BL Lacs versus BCUs.   Similar variability between BL Lacs and BCUs implies that these sources share more common features, making it challenging to distinguish them solely based on variability metrics.
These findings align with earlier studies \citep{2011ApJ...743..171A,2010ApJ...716...30A} that indicate FSRQs tend to exhibit more pronounced variability due to more powerful jets and accretion processes.  It is known that FSRQs occupy denser and more gas-rich environments than BL\,Lac objects, which in turn results in higher liminosity and variability. For example, \citet{2011ApJ...736..128P} noted that FSRQs feature the strong broad emission lines a clear evidence of abundant circumnuclear gas  unlike in BL\,Lacs. On average, FSRQs are much more luminous than BL\,Lacs. In $\gamma$-rays, for example, \citet{2020A&A...634A..80R} show that FSRQs cluster at much higher luminosities than BL\,Lacs by using the  sample of Fermi blazars.
The greater luminosity of FSRQs is also evident in emission lines: strong broad lines and dusty torus emission in FSRQs indicate more accretion power and richer photon fields e.g. \citep{2011MNRAS.414.2674G}. By contrast, BL\,Lacs often show only weak or no lines, consistent with low accretion luminosity. Therefore, the observation that a substantial fraction of BCUs exhibit variability characteristics closely resembling those of BL\,Lacs suggests that many BCUs may also inhabit similarly low-density environments.

In our study, we considered full light curves including the flaring and low flux states. We know a large fraction of blazars exhibit variability dominated by occasional strong flares separated by prolonged quiescent states. The variance statistic captures the total variability in the light curve, meaning that a few intense flares can contribute significantly to the overall variance. To examine this, one could analyze the light curves by separating flaring and quiescent states or use additional measures such as the fraction of total variance contributed by the highest flux states. In our analysis, the variability statistic does not explicitly distinguish between variability from frequent moderate fluctuations and variability dominated by rare, intense flares. However, one way to address this would be to compute the variance separately for different flux levels or analyze the light curves.

 The analysis of the correlation between $\gamma$-ray flux and spectral index in blazars reveals a moderate positive correlation across different time binnings (3-day, 7-day, and 30-day) for BL Lacs and BCUs, indicating that they  tend to become spectrally softer as their flux increases. This ``softer when brighter" behavior contrasts with the commonly observed "harder when brighter" trend seen during short flaring events, where higher fluxes are typically associated with harder spectra \citep{2010ApJ...716...30A,2011A&A...530A..77F}. 
Here, more particles may be injected with a harder spectrum (if acceleration is efficient), which would make the spectrum harder. But if the cooling is more efficient, the high-energy particles lose energy quickly, leading to a softer spectrum. So  a rapid injection of particles and then quickly cooling lead to a steeper spectrum.
This ``softer when brighter" trend also suggests that the underlying physical processes governing long-term $\gamma$-ray emission in BL\,Lacs may involve shifts in particle acceleration or cooling mechanisms over time. 
 In contrast, FSRQs exhibit a mild anticorrelation (see Table \ref{tab:spe}), suggesting a tendency for these sources to become spectrally harder as their flux increases. Statistical analysis shows that the P-values for the 7-day and 30-day bins reject the null hypothesis of no correlation between flux and index, whereas for the 3-day bin, the null hypothesis cannot be rejected. The absence of correlation in the 3-day bin suggests that short-term variability may obscure any consistent flux-index relationship in FSRQs. 
This timescale dependence highlights a critical distinction: long-term trends reflect persistent physical mechanisms (e.g., sustained particle injection or gradual cooling), whereas short-term variability may arise from localized, transient phenomena like shocks or magnetic reconnection \citep{2009MNRAS.397..985G}. These findings imply that the environment near the blazar’s central engine evolves over longer periods, influenced by factors such as magnetic field strength, particle density, and energy dissipation rates, which shape the spectral characteristics of the emission. The need for multi-timescale observations becomes evident, as studies focusing solely on short-term variability may overlook the broader trends governing blazar behavior over time.
 These results  highlight the importance of integrating data across different timescales to achieve a comprehensive understanding of these objects, as emphasized in previous works \citep{Ackermann_2011,2016ARA&A..54..725M}.
 
  Our statistical study both variance and index/flux correlation shows that a large fraction of BCUs eventually gets classified as BL\,Lacs.
This aligns with previous studies that have found BCUs to exhibit similar non-thermal emission characteristics and variability amplitudes as BL Lacs (\citep{2020Ap&SS.365...12D, 2016Ap&SS.361..316A, 2016MNRAS.462.3180C}). In Figure \ref{fig:scatter_mean_index_flux}, where the mean index is plotted against mean flux,  BL\,Lac sources tend to show larger scatter than FSRQs. This is expected, since BL\,Lac sources are composed of three subclasses, HBL, IBL and LBLs. Importantly, the results of variability analysis, which shows that BCUs are more likely to be BL\,Lac are consistant with the flux vs index plot (Figure \ref{fig:scatter_mean_index_flux}) as both BL\,Lac and BCU shows mild brighter softer trend while FSRQs shows slight harder brighter trend. The narrow spread of BCUs suggest that majority of them may belong to one subclass of BL\,Lacs. These results are consistent with the \citep{2016MNRAS.462.3180C} which used flare light-curve patterns  and neural networks on 3FGL data and assign majority of BCUs to BL\,Lac class. 
Studies such as \citet{2017A&A...602A..86L} applied machine-learning classifiers to Fermi BCUs: they found that $\sim 91\%$ of their BCU sample were classified as BL Lac-like while only $\sim 3\%$ were FSRQ-like.
\citep{2020Ap&SS.365...12D} in their optical spectroscopic observations of 37 sources, mostly BCUs, they confirmed the BL\,Lac nature of 27 sources. This finding supports the idea that a significant fraction of BCUs are intrinsically BL\,Lacs. 

 The examination of long-term flux distributions in blazars provides key insights into the physical processes driving their variability.   It is well established that purely additive fluctuations, arising from the linear sum of numerous independent, small-amplitude variations, lead to normally distributed total fluxes, as expected from the central limit theorem \citep{2003MNRAS.345.1271V, 2012MNRAS.421.2854S}. However, observational evidence across various accreting systems consistently reveals lognormal flux distributions, implying that the underlying variability mechanism is multiplicative in nature \citep{1997MNRAS.292..679L, 2005MNRAS.359..345U}. In accretion-powered sources like X-ray binaries and AGNs, this behavior is well explained by the 'propagating fluctuations' model, where local variations in the mass accretion rate generated at different disc radii propagate inward, coupling multiplicatively and modulating the emission from the inner regions \citep{1997MNRAS.292..679L}. Interestingly, lognormality is also observed in non-thermal jet-dominated systems such as blazars \citep{2018RAA....18..141S, 2020MNRAS.491.1934K, 2020MNRAS.496.3348S}, possibly due to the imprint of disc fluctuations onto the jet via efficient variability transmission mechanisms \citep{2010LNP...794..203M, 2019MNRAS.482.2447P, 2022MNRAS.510.3688K}. Nevertheless, observations of extremely rapid, high-energy variability suggest that jet emission can at times evolve independently of the disc \citep{1996Natur.383..319G, 2007ApJ...664L..71A, 2012MNRAS.420..604N}. In such scenarios, lognormal behavior may arise from intrinsic jet processes, such as linear Gaussian perturbations to particle acceleration or escape timescales \citep{2018MNRAS.480L.116S}, or even from additive processes like shot noise when modified by Doppler boosting from a population of randomly oriented mini-jets \citep{2012A&A...548A.123B}.

 In our analysis, $\gamma$-ray flux distributions for blazars were studied across different time bins (3-day, 7-day, and 30-day) using the AD test. The AD test results confirmed log-normality in a significant number of BL Lacs, FSRQs, and BCUs, indicating that these systems exhibit multiplicative variability processes. Flux points with large errors were excluded to maintain data reliability, and outliers were filtered to focus on the general behavior of the systems. The absence of strong correlations between the $\mu$ and $\sigma$ of the flux distribution across different time bins suggests that flux variations are driven by complex processes that do not depend solely on the average flux values.  In the 30-day binned light curves, BCUs show a tendency to behave similarly to FSRQs, implying that longer observation periods may reveal  patterns in their variability.

For sources where the log-normality of flux distributions was rejected, a bimodal PDF was fitted, indicating that some blazars may experience two distinct states of variability. Positive correlations between the centroids of the bimodal distributions in all blazars suggest that when flux increases in one state, it tends to rise in the other as well, reflecting possible transitions between different physical states of activity. In contrast, the negative correlation between the widths of the distributions ($\sigma_1$,  and $\sigma_2$ ), particularly for BL\,Lacs, points to variability behaviors, where the spread in flux decreases as the system enters a different state. Additionally, the analysis shows that FSRQs exhibit consistent positive correlations across multiple parameters, implying more structured variability.  The results indicate that blazars, particularly BCUs, behave differently depending on the time scale of observation, with longer bins showing stronger correlations and clearer patterns of variability. 
In this study,  many blazars retain lognormal behavior in both 3-day and 7-day binned light curves, indicating that the multiplicative processes responsible for this behavior are persistent across these timescales. However, some sources only show log-normality in one binning scheme, which might indicate a transition between different variability modes. The observation of double lognormal flux distributions in some sources suggests that there could be two distinct multiplicative processes influencing the variability. 

The PSD is calculated to study the variability patterns in blazars. 
Typically, the slope of the PSD provides insights into the nature of variability. A PSD slope close to 0 indicates white noise (random variability), while steeper slopes suggest more structured, long-term variability. The PSD slopes for BL Lacs are typically less than 1 in both 3-day binned light curves. The PSD slopes in FSRQs tend to be steeper than those in BL Lacs, especially in the 7-day binned light curves. This suggests that the variability in FSRQs is not dominated by white noise but follows a more structured, red noise pattern (with longer timescales contributing more to variability).  Notably, BL Lacs with PSD slopes consistent with white noise (e.g.,4FGL 1806.8+6949, slope 0.10$\pm$0.13) may represent systems where short-timescale variability dominates, masking underlying trends.
 In our study, PSD slopes for BL\,Lac objects range from 0.1 to 0.83 for 3-day binned light curves and from 0.02 to 1.40 for 7-day bins. For FSRQs, the PSD slopes range from 0.10 to 0.93 (3-day bins) and 0.56 to 1.45 (7-day bins).
These values are generally flatter than those reported in early studies such as \citet{2010ApJ...722..520A}, who analyzed 11 months of 7-day binned light curves for bright blazars and found most PSD slopes in the range 1.1 to 1.6. Similarly, \citep{2012ApJ...749..191C} reported R-band optical PSD slopes ranging from 0.6 to 2.3, with an average slope of $- 1.6 \pm 0.3$ across six blazars. However, they also noted significant variation between optical and $\gamma$-ray PSDs—for example, PKS 2155−304 showed a steep optical PSD ($\sim -2.2$) but a much flatter $\gamma$-ray PSD \citep{2010ApJ...722..520A}, highlighting wavelength-dependent variability behavior.
More recent studies, however, have reported flatter $\gamma$-ray PSD slopes consistent with our findings. \citet{2014ApJ...786..143S} examined 4-year Fermi-LAT light curves of 13 bright blazars and found slopes typically $\leq 1$, with no systematic difference between FSRQs and BL Lacs. \citet{2013ApJ...773..177N} reported a slope of $0.38 \pm 0.21$ for Mrk 421, which is even flatter than the $0.69 \pm 0.07$ we obtained for the same source. \citet{2017ApJ...849..138K} analyzed 7-year Fermi-LAT data for Mrk 421, B2 1520+31, and PKS 1510−089, and their results for PKS 1510−089 agree with those from \citet{2020ApJS..250....1T}. Similarly, \citet{2017MNRAS.471.3036P} reported PSD slopes between 0.5 and 0.8 for their sample.
\citet{2019ApJ...877...39M} investigated 9.5 years of weekly binned Fermi-LAT data for six FSRQs and found flatter PSD slopes in the range 0.65 to 1.1. Among the sources highlighted here only  sources, 3C 279 and Mrk 421 are also present in our sample, for which we obtained flater slopes of $0.80 \pm 0.08$ and $0.69 \pm 0.07$, respectively.
These comparisons suggest that methodological differences—including the PSD estimation technique (e.g., Fourier transform vs. Lomb-Scargle periodogram), binning cadence, and the duration and brightness of light curves—can significantly influence the estimated PSD slopes. Importantly, our use of longer-duration data, including both high- and low-brightness phases and a broader sample of blazars, naturally leads to a wider distribution of slope values.

Binning the data can influence the PSD slope and the distribution of flux values. In our analysis, some sources show log-normality in the 3-day binned light curves but not in the 7-day binning (and vice versa). This suggests that the variability mechanisms may operate on different timescales, and binning might highlight or obscure certain behaviors. Binning can smooth out shorter-term variations, which might explain the disappearance of log-normality in certain cases when transitioning to a longer bin. 
Additionally, the lack of PSD break implies no observed characteristic timescale within the {\emph Fermi} observational window, consistent with a superposition of stochastic events (e.g., shock-in-jet models). Breaks are usually associated with physical processes like cooling or crossing a specific size scale in the jet or emission region. The lack of such breaks indicates continuous variability without a dominant timescale. 

The spectral index distribution analysis of BL\,Lac and FSRQ sources over different binning intervals reveals the complexity of the underlying variability processes in these sources. While normality was accepted in select cases, the prevalence of log-normal/double log-normal fits suggests a more intricate spectral index distribution. This observation aligns with the idea that blazar variability is often non-linear and multi-model, which is consistent with other studies showing that blazars exhibit flux variations across different timescales driven by stochastic processes in the relativistic jets. The higher reduced-$\chi^2$ values for sources like 4FGL\,1104.4+3812 and 4FGL\,2158.8-3013  indicate that even double log-normal models may not capture all the variability, hinting at the possible presence of additional physical mechanisms or more complex emission models that govern the spectral behavior of these sources. This calls for the application of more sophisticated statistical models to better characterize the spectral index distributions, particularly when examining longer-timescale variability in blazars.
Understanding the longer variability patterns  is crucial for refining theoretical models of blazar emission and improving predictions of their variability patterns across different energy bands.

\section{Conclusion}
\label{sec:conclusion}

 This comprehensive study of long-term \emph{Fermi}-LAT $\gamma$-ray light curves for a large sample of blazars (FSRQs, BL Lacs, and BCUs) across multiple time binnings (3-day, 7-day, and 30-day) reveals that variability characteristics depend strongly on both blazar subclass and observational timescale. FSRQs consistently show higher fractional variability ($F_{\rm var}$) than BL Lacs and BCUs, supporting the view that their powerful jets and high accretion rates drive stronger flux variations. BCUs display intermediate behavior, often resembling BL Lacs in variability amplitude. The correlation between $\gamma$-ray flux and spectral index shows that BL Lacs and BCUs generally follow a “softer when brighter” trend, indicative of cooling-dominated processes, while FSRQs exhibit a slight “harder when brighter” trend, especially at longer time bins, hinting at different acceleration or emission mechanisms. The lack of flux–index correlation in FSRQs at shorter time bins suggests that short-term flares may obscure underlying long-term processes. Lognormal flux distributions, observed in many sources across all time bins, point to multiplicative variability processes likely tied to turbulent accretion flows or jet dynamics. Their persistence across timescales suggests such processes operate over a broad temporal range. The appearance of bimodal or double-lognormal distributions in some blazars indicates distinct variability states or multiple overlapping multiplicative processes. Positive correlations between bimodal centroids imply coherent flux changes across states, while negative correlations between distribution widths—especially in BL Lacs—suggest reduced variability during specific activity phases. FSRQs, in contrast, show consistent positive correlations across multiple parameters, indicating more structured variability. PSD analysis reveals a wide diversity of slopes, reflecting variability across a broad range of timescales. BL Lacs typically show flatter PSD slopes ($<1$), particularly in 3-day bins, while FSRQs tend to exhibit steeper slopes--especially in 7-day bins--indicating more structured, red-noise-like behavior. The flatter slopes reported here, compared to earlier studies, align with more recent long-term monitoring and suggest that variability characteristics evolve with extended observations. Differences across studies likely arise from variations in binning cadence, PSD estimation methods, and inclusion of diverse brightness states. Overall, these results highlight the importance of long-duration, multi-timescale monitoring to unravel the physical mechanisms driving blazar variability and to better understand their jet dynamics and central engines.

\section{Acknowledgements}
 ZS is supported by the Department of Science and Technology, Govt. of India, under the INSPIRE Faculty grant (DST/INSPIRE/04/2020/002319). We express gratitude to the {\emph Fermi}-LAT for providing  a comprehensive long-term $\gamma$-ray data of high-energy sources.

\section{Data Availability}
This research has used $\gamma$-ray light curves from Fermi LCR \citep{2023ApJS..265...31A} which can be accessed at https://fermi.gsfc.nasa.gov/ssc/data/access/lat/LightCurveRepository/ .\\

\bibliographystyle{apsrev4-2} 
\bibliography{apssamp}

\end{document}